\documentclass{aa}  
\usepackage{graphicx}
\usepackage{natbib} \bibpunct{(}{)}{;}{a}{}{,} % to follow the A&A style
\usepackage{txfonts}
\usepackage{color}
\usepackage{multirow}
\usepackage{orcidlink}

\def\deg{\ifmmode{^\circ}\else$^{\circ}$\fi}	              % degree
\def\kms{\ifmmode {{\rm \;km\;s^{-1}}}		    	      % km s-1
       \else {\hbox{$\,${\rm km$\;$s$^{\rm -1}$}}}\fi}
\begin{document} 

   \title{CII fine-structure line observations of the Sagittarius C Region in the Galaxy's Central Molecular Zone}

    \author{D. Riquelme-V\'asquez\orcidlink{0000-0001-5389-0535}\inst{1,2} 
    \and R. G\"usten\orcidlink{0000-0002-1708-9289}\inst{1} 
    \and M.R. Morris\orcidlink{0000-0002-6753-2066}\inst{3} 
    \and A.I. Harris\orcidlink{0000-0001-6159-9174}\inst{4} 
    \and M.A. Requena-Torres \inst{5}
    \and E.F.E. Morales \inst{6}
    \and J.~Stutzki\orcidlink{0000-0001-7658-4397}\inst{7} 
    \and R.~Simon\orcidlink{0000-0003-2555-4408}\inst{7}  
    \and C.~Risacher\orcidlink{0000-0001-7162-4185}\inst{8}
    \and R. Higgins \inst{7}
}
    \institute{Max-Planck-Institut f\"ur Radioastronomie, Auf dem H\"ugel 69, 53121 Bonn, Germany
    \and Departamento de Astronom\'ia, Universidad de La Serena, Ra\'ul Bitr\'an 1305, La Serena, Chile\\
    \email{denise.riquelme@userena.cl} 
    \and Department of Physics and Astronomy, University of California, Los Angeles, CA 90095, USA
    \and Department of Astronomy, University of Maryland, College Park, MD 20742, USA
    \and Department of Physics, Astronomy, and Geosciences, Towson University, Towson, MD 21252, USA
    \and Independent researcher
    \and I. Physikalisches Institut der Universit\"at zu K\"oln, Germany   
    \and Institut de Radioastronomie Millim\`{e}trique, 300 rue de la Piscine, Domaine Universitaire, 38406 Saint Martin d'H\`{e}res, France
      }
   \date{Received ; accepted }
\abstract
   {Sagittarius C (Sgr C) is a massive but relatively quiescent complex at the western edge of the Galaxy's Central Molecular zone (CMZ). While the Sgr B2 region on the opposite side of the CMZ has been extensively studied, Sgr C has received comparatively less attention.}
   {We aim to characterize the kinematics and physical state of the atomic gas in the Sgr C region using spatially and velocity-resolved emission from the [CII] line. This line traces the multi-phase gas consisting of the warm ionized medium, the warm and cold diffuse atomic medium, and the warm dense molecular gas, providing a complement to molecular line, dust, infrared, and radio observations.}
   {We present a fully sampled 74 $\times$ 47 pc map of the [CII] 158 $\mu$m fine-structure line toward Sgr C, observed with the upGREAT receiver onboard the Stratospheric Observatory for Infrared Astronomy (SOFIA) airborne observatory. The data feature a spatial resolution of 0.55 pc and a spectral resolution of 1 \kms. These observations are analyzed in conjunction with ancillary maps of the $J=2-1$ transition of CO and its isotopologues observed with the PI230 receiver at the Atacama Pathfinder Experiment (APEX) telescope.}
   {[CII] emission is widespread throughout the region, showing a continuous structure extending from Sgr A to Sgr C with complex morphology. The bulk of the emission arises from gas at negative radial velocities, consistent with the sense of Galactic rotation. The most prominent feature is the giant Sgr C HII region, where [CII] reveals an expanding, ring-like shell structure interpreted as a photo-dissociation region (PDR). We modelled the shell's kinematics, deriving an expansion velocity of $\sim 23$ \kms\ and a dynamical age of $\sim 0.13$ Myr. Our analysis suggests that stellar winds from the known massive stars are likely insufficient to power the observed expansion, pointing toward alternative driving mechanisms such as a buried supernova. We find a striking spatial association between this shell and a non-thermal radio filament, providing evidence that the shell's expansion into the surrounding medium has triggered high-mass star formation at its edge.}
  {}
  
   \titlerunning{CII fine-structure line observations of the Sagittarius C Region in the CMZ}
   \maketitle
\nolinenumbers 
%%-----------------------------------------------------------------  
\section{Introduction} \label{intro}
Studying the Galactic center (GC) of our own Galaxy offers a unique opportunity to explore and understand, by analogy, the structure and astrophysics of external galactic nuclei. The central $\sim500$ parsecs of our Galaxy contain a huge concentration of molecular gas known as the Central Molecular Zone \citep[CMZ, ][]{Morris_Serabyn_1996} with extreme and peculiar physical and chemical properties, such as high kinetic temperatures \citep{Guesten_et_al_1981, Huettemeister_et_al_1993b, Ginsburg_et_al_2016}, strong and widespread emission from several shock tracers \citep[e.g., SiO;][]{Martin-Pintado_et_al_1997, Huettemeister_et_al_1998, Riquelme_et_al_2010b}, and large linewidth that indicate strong turbulence that leads to some suppression of star formation \citep{Longmore_et_al_2013a}.

The Sgr C complex (G359.45--0.05) offers a unique opportunity to provide further insight into the CMZ. This massive molecular complex is located on the negative-longitude side of the CMZ (extending from Galactic longitudes of $359.4^\circ$ to $359.6^\circ$, which corresponds to $\sim 29$~pc; \citealt{Bally_et_al_1988}), and is situated $\sim 30'$ (75~pc) in projection from the Sgr A complex. This region, Sgr C, contains less molecular material overall, as roughly 3/4 of the $^{13}$CO (1--0) emission in the CMZ is found at positive Galactic longitudes \citep{Bally_et_al_1988}.

Sgr C reveals a wide variety of physical phenomena, such as one of the brightest extended HII regions in the GC (hereafter the Sgr C HII region, powered by a central O4 star and at least partially by a late Wolf-Rayet star; \citealt{Clark_et_al_2021}). The thermally emitting HII region is close, at least in projection, to a bright nonthermal radio filament \citep{Liszt_Spiker_1995, Lang_et_al_2010} and a high-mass star-formation region containing outflows and high-velocity gas with forbidden Local Standard of Rest (LSR) velocities relative to the sense of Galactic rotation (i.e., positive velocities at negative Galactic longitudes). These properties make it a critical location to study star formation and its efficiency within the CMZ. The physical properties of the molecular cloud are comparable to those of other remarkable non-star-forming clouds such as G0.253+0.016. Even with a mass of $\sim 10^5$ M$_{\odot}$ of gas and dust, with temperatures and column densities favorable for massive star formation to occur, the Sgr C molecular cloud only shows star formation toward the tip of the cloud (the cusped edge of the molecular cloud at its extreme western extent; see, e.g., Fig. 4 in \citet{Kendrew_et_al_2013}).

Ongoing high-mass star formation in Sgr~C was reported by \citet{Forster_Caswell_2000}, who detected an ultracompact HII region ($1.4'' \times 0.3''$, or $0.07 \times 0.01$~pc$^2$ at an assumed distance of 10~kpc) associated with an H$_2$O maser. \citet{Yusef-Zadeh_et_al_2009} detected a region of extended 4.5~$\mu$m emission within $5''$ of the ultracompact radio source, a so-called Extended Green Object \citep[EGO;][]{Cyganowski_et_al_2008, Chambers_et_al_2009}, which is strongly linked to early-stage high-mass star formation and associated outflows. \citet{Kendrew_et_al_2013} confirmed the early stages of high-mass star formation with the detection of two protostellar cores and several knots of H$_2$ and Brackett-$\gamma$ emission associated with the compact radio source, indicative of an outflow. However, there is no evidence for ongoing star formation outside the EGO, which indicates that most of the cloud is quiescent. Recently, \citet{Crowe_et_al_2025} revealed the nature of this source in more detail; using the James Webb Space Telescope (JWST), they identified G359.44--0.102 as a protocluster embedded within the main Sgr~C molecular cloud, further characterizing the massive protostars and their outflows within it.

Eight arcminutes north of the Sgr~C complex, \citet{Yusef-Zadeh_et_al_2009} found a cluster of 24~$\mu$m sources distributed toward a bright nonthermal radio filament, which they interpreted as 18 Young Stellar Objects (YSOs) with individual masses in the range of 5--11.7~$M_{\odot}$ and luminosities between $6.5 \times 10^2$ and $10^4$~$L_{\odot}$. Nogueras-Lara (priv. comm.) found that the immediate vicinity of the Sgr~C HII region hosts several $10^5$~$M_{\odot}$ of young stars with ages of at least $\sim 20$~Myr, though this older population is distinct from the source powering the younger expanding HII region.

Observing the fine-structure line emission of ionized carbon is inarguably the most important far-IR spectral tool for understanding the state and dynamics of the interstellar gas in galactic nuclei. With an energy above the ground state of 91~K, the upper level of the 158~$\mu$m [CII] transition is readily excited by collisions with electrons and neutral species, typically making it the dominant cooling line of the interstellar medium (ISM) \citep{Hollenbach_Tielens_1999}.

In all but the most extreme galactic nuclei, the 158~$\mu$m [CII] line accounts for 0.1\% to 1\% of the bolometric luminosity of the entire galaxy \citep{Stacey_et_al_1991, Malhotra_et_al_2001, Gracia-Carpio_et_al_2011}. [CII] traces the warm ionized medium, the warm and cold diffuse atomic medium, and the warm dense molecular gas. Recent observations \citep{Pineda_et_al_2013_I, Langer_et_al_2014_II, Pineda_et_al_2014_III, Langer_et_al_2014, Langer_et_al_2017, Perez-Beaupuits_et_al_2015} have firmly established that the line traces both the ionized gas and the neutral gas exposed to UV photons from associated high-mass stars.

[CII] therefore provides a measure of the dense atomic gas (photodissociation regions, with typical densities of $n \sim 10^3$--$10^6$~cm$^{-3}$; \citealt{Mookerjea_et_al_2021}) in addition to ionized and diffuse atomic components, and complements the existing atomic, molecular, and dust data on the CMZ.

%--------------------------------------------------------------------
\section{Observations and data reduction}

\subsection{SOFIA Observations}
The observations are part of a large-scale survey of the CMZ in [CII] (G\"usten et al., in prep.). Part of the data, toward the Sgr~A and Sgr~B regions, has been published previously \citep{Harris_et_al_2021, Harris_et_al_2025}.

The observations of the Sgr~C region were carried out using the upgraded German Receiver for Astronomy at Terahertz Frequencies \citep[upGREAT;][]{Risacher_et_al_2016, Risacher_et_al_2018} onboard the Stratospheric Observatory for Infrared Astronomy (SOFIA; \citealt{Young_et_al_2012}) under project numbers 06\_0157 and 83\_0609 during the 2018 southern deployment to Christchurch, New Zealand, on June 5, 6, 7, 18, 20, 21, and 22. The observations were taken at altitudes between 11.5 and 13.1~km. We observed the $^2$P$_{3/2}$--$^2$P$_{1/2}$ transition of [CII] at 1.9005369~THz ($\lambda 157.74~\mu$m) using the $2 \times 7$-pixel dual-polarization array (the horizontal polarization of the array is referred to as LFAH, while its vertical polarization is referred to as LFAV). We used the Fast Fourier Transform Spectrometers (FFTS4G), an updated version of the instrument described by \citet{Klein_et_al_2012}. The spectrometers provide 32\,768 channels per pixel across the signal and image bands, resulting in 16\,384 channels per spectrum for the signal band. The frequency resolution is 0.2441~MHz, which corresponds to a native velocity resolution of 0.0385~\kms\ at 1.9~THz. While the observations cover a velocity range of $-286$ to $+344$~\kms, we restricted the final dataset to the range of $-200$ to $+344$~\kms\ to avoid contamination from atmospheric features at lower velocities.

The Sgr~C complex was covered with a mosaic of $3 \times 2$ tiles of $609'' \times 567''$ each, with an overlap of two rows/columns. Each tile was mapped in the on-the-fly (OTF) observation mode \citep{Mangum_et_al_2007} with a step size of $7''$, a sampling of 0.3~sec/dump, and $7''$ separation between rows. To optimize observing efficiency under the observational constraints of this airborne observatory (cavity opening), we utilized a two-step off-position strategy. A relatively nearby position (J2000 $17^{\text{h}}44^{\text{m}}14.21^{\text{s}}$, $-28^\circ11'17.5''$) was used as the reference off-position for the OTF maps. This position was subsequently observed against a more distant, clean position (J2000 $17^{\text{h}}41^{\text{m}}55.21^{\text{s}}$, $-28^\circ23'33.3''$) to identify contamination in the near-off. Three and a half of the six tiles were observed twice, in orthogonal scanning directions, to increase the signal-to-noise ratio and reduce residual scanning patterns.  

The calibration was performed using the KOSMA atmospheric calibration software for SOFIA/GREAT (\citealt{Guan_et_al_2012}, version May 2018). For the opacity correction, the precipitable water vapor column was obtained from a free fit to the atmospheric total power emission spectrum. The dry constituents were fixed to the standard model values. All receiver and system temperatures are given on the single-sideband scale. The antenna temperature ($T^*_A$) was converted to $T_{\text{mb}}$ using the beam efficiencies that were computed individually for each pixel and polarization ($\langle \eta_{\text{mb}} \rangle = 0.65$). The main beam size is $14.1''$. We estimate that the uncertainty in the amplitude calibration is less than 20\%.

\subsection{APEX Observations}
Complementary observations of the molecular gas toward Sgr~C were obtained using the Atacama Pathfinder Experiment (APEX) telescope \citep{Gusten_et_al_2006}. We used the two-sideband, dual-polarization PI230 heterodyne receiver to map the $J=2-1$ transitions of $^{12}$CO, $^{13}$CO, and C$^{18}$O. The receiver was connected to the Fast Fourier Transform Spectrometer (FFTS4G) \citep{Klein_et_al_2012}, with a frequency resolution of 61.03~kHz (corresponding to $0.079$--$0.083$~\kms). The beam sizes for $^{12}$CO, $^{13}$CO, and C$^{18}$O are $28.7''$, $30.1''$, and $30.2''$, respectively. These data were observed as part of a large survey covering the whole CMZ (Riquelme-V\'asquez et al., in prep.), and all observation and data reduction details will be described there.

\subsection{Data Reduction}
Given the large linewidths, with multiple broad components blending along the line of sight (typically observed in Galactic center clouds) and the limited available baseline for negative velocities due to atmospheric features, the data reduction posed significant challenges. A conventional polynomial approach to removing baselines was insufficient. Additionally, the data exhibited broad baseline features likely caused by gain instabilities \citep[see Section 3.3 in][]{Higgins_et_al_2021}, which could affect any of the 14 pixels. Some of the data also suffer from radio interference (RFI) \citep[see Section 3.4 in][for details]{Higgins_et_al_2021}, which fortunately only affected pixel LFAH-3. Accordingly, all data from pixel LFAH-3 were discarded to maintain data integrity.

For this reason, the data reduction was performed in three steps: initial baseline correction using the scaled spline method, standard polynomial baseline subtraction with the CLASS package, and a final cleaning considering the statistics of the reduction. The normalized distributions of the RMS for the LFAH and LFAV polarizations in Fig.~3 (top and middle panels) are shown to characterize the receiver performance across the arrays before the final data processing steps.

First, we used the "scaled spline method" \citep[see Section 3.2 in][for details]{Higgins_et_al_2021} to correct the baseline features in the calibrated data. This technique was developed by \citet{Kester_et_al_2014} and \citet{Higgins_2011} (and was already successfully used for the Sgr~B [CII] data reduction in \citealp{Harris_et_al_2021}). This technique fits a spline to the spectra generated from the difference between OFF observations to generate a catalog of baseline shapes, which are then scaled and applied to correct the ON--OFF astronomical data.

After the initial baseline correction using the scaled spline method, the data were further reduced using the Continuum and Line Analysis Single-dish Software (CLASS) package from the GILDAS software\footnote{\url{https://www.iram.fr/IRAMFR/GILDAS/}}. All spectra were resampled to 1~\kms. Due to the large linewidths observed across the entire Sgr~C region, each tile was individually reduced, carefully selecting the velocity range for baseline subtraction. A first-order polynomial base was used if the root mean square (RMS) noise of the off-line spectral channels of each spectrum was less than 1.5 times the theoretical noise. The use of a higher-order polynomial for baseline subtraction was not feasible due to the large linewidth and restricted baseline range available, which would introduce spurious features into the data.
The theoretical noise ($\sigma_{\text{theo}}$) for the upGREAT receiver is computed according to the radiometer equation \citep{Tools}.

Theoretical noise values range between 2.6 and 9.3~K with an average of 4.2~K. Data with higher RMS values were not used, as they did not exhibit a Gaussian distribution in the RMS histogram. With this filter, approximately 99.2\% of the total spectra were retained.

\begin{figure}
 \centering
\includegraphics[width=1.0\hsize, angle=0]{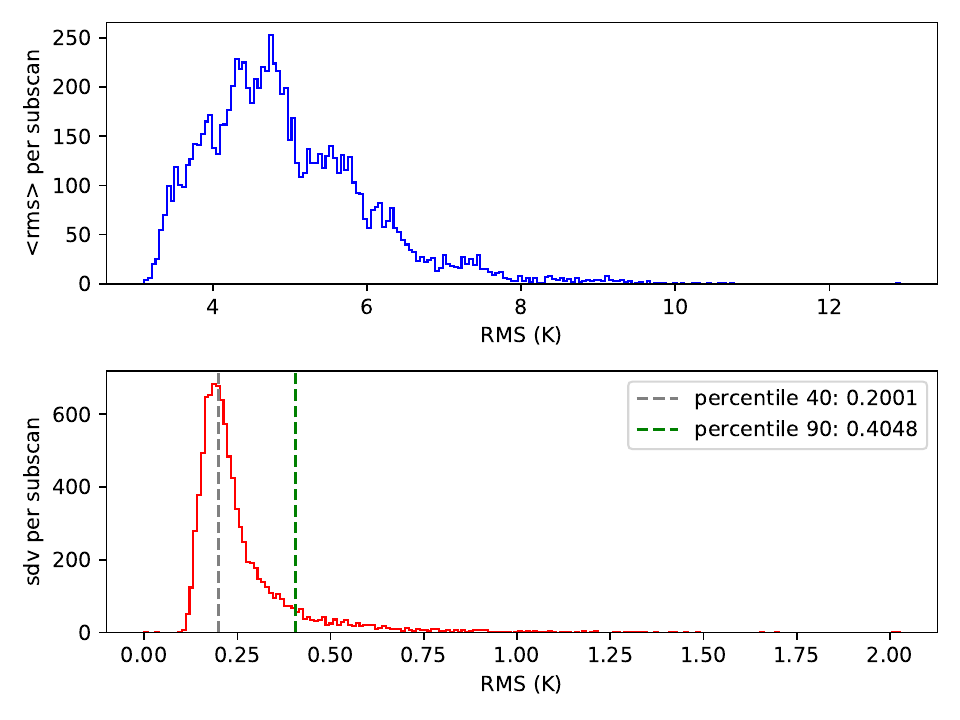}
\caption{Average of the spectral RMS noise computed for each subscan for the Sgr C observed region (top) and standard deviation (sdv) computed for each subscan (bottom). Data are resampled to 1 \kms\ resolution, and the velocity range for baseline subtraction was carefully selected considering the emission-free channels for each observed tile. See text for details.}
\label{percentiles}
\end{figure}

Because the data still had some spurious features, we proceeded to inspect and discard data using the statistics of the reduction over smaller regions. First, we computed the average RMS noise for each subscan for the entire survey, along with its standard deviation (Fig.~\ref{percentiles}). Subscans with a standard deviation lower than the $40^{\text{th}}$ percentile of the dataset were retained. Subscans with a standard deviation higher than the $40^{\text{th}}$ percentile but lower than the $90^{\text{th}}$ percentile only retained data with an individual dump RMS noise lower than the average of the subscan plus three times the standard deviation of the OTF line RMS noise. For subscans with a standard deviation higher than the $90^{\text{th}}$ percentile, data were kept only if the RMS noise of each spectrum was lower than the average of the RMS of its subscan. This filtering process was successful in removing nearly every bad spectrum from the survey, retaining 95.0\% of the data.

Even after applying this filter, some remaining bad baseline features were present that could not be corrected using a higher-order polynomial. To discard these remaining bad spectra, the average integrated intensity was computed for each tile within the velocity range most affected by this issue ($-180$ to $-90$~\kms). For each spectrum within a tile, if the integrated intensity fell between the average plus or minus two times the standard deviation of the tile, the spectrum was retained; otherwise, it was discarded. For the two tiles covering the northeast part of the map, which were mapped in only one scanning direction, a more restrictive criterion was applied to mitigate striping features in the cube. Only data with an RMS noise lower than the average RMS plus or minus one standard deviation were retained.

\begin{figure}
\centering
\includegraphics[width=1.0\hsize, angle=0]{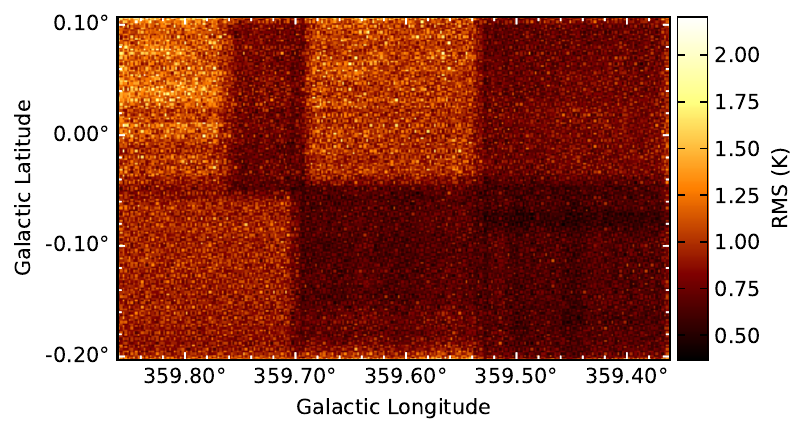}
\caption{Distribution of the RMS over the final map (after regridding and converted to Galactic coordinates). The RMS is computed over the line-free velocity range from 150 to 250 \kms\ in 1 \kms\ bins. Those tiles that were observed in only one scanning direction show increased noise.} 
\label{noisemap}
\end{figure}

\begin{figure}
\centering
\includegraphics[width=1.0\hsize, angle=0]{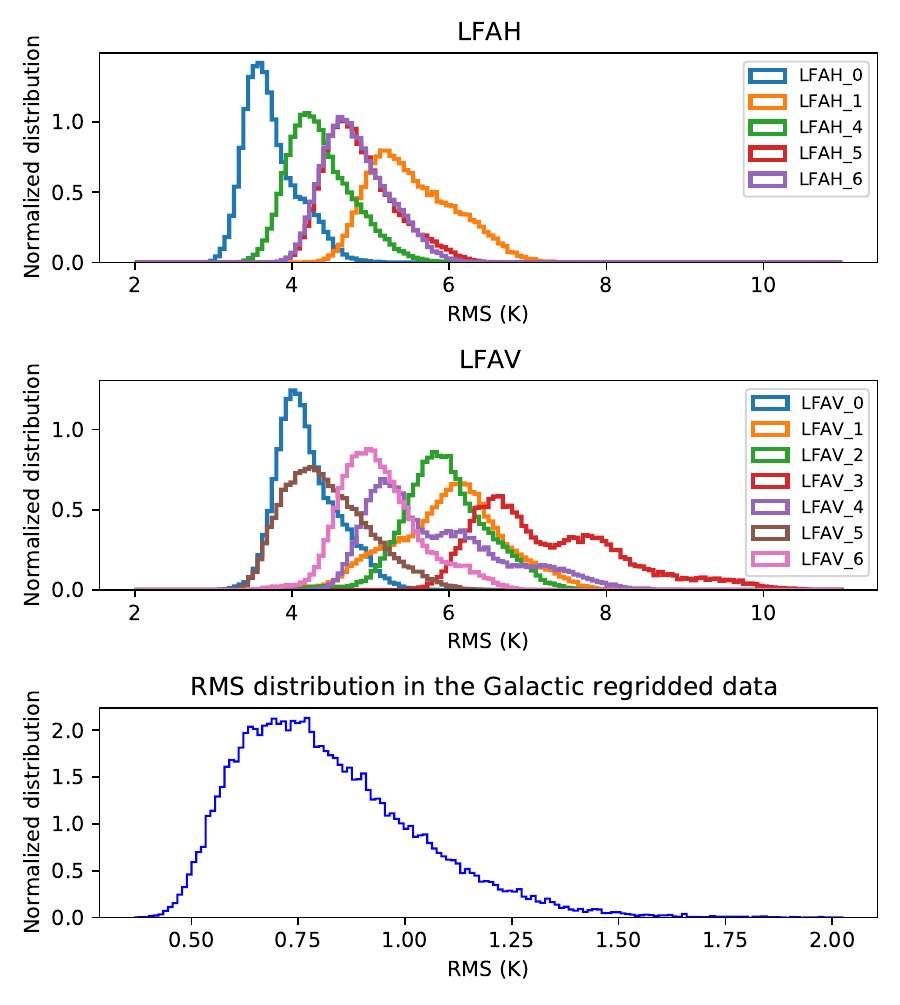}
\caption{Normalized distribution of the RMS per pixel for the LFAH polarization (Top) and LFAV polarization (Middle). Bottom: Normalized distribution of the RMS in the regridded data converted to Galactic coordinates (final data cube, the convolved spectra in the final data cube). The average of the RMS of the final data cube is 0.82 K (median 0.78 K), and the standard deviation is 0.21 K.}
\label{histonoise}
\end{figure}

With this filtering, we retained 84.8\% of the total data and successfully removed most of the striped features in the final data cube. 
The data were then regridded in equatorial coordinates weighted by their RMS and subsequently converted to Galactic coordinates using standard CLASS routines. Figure~\ref{noisemap} displays the RMS noise map of the final dataset converted to Galactic coordinates, while Figure~\ref{histonoise} presents the histogram of the RMS values for the complete dataset before regridding (top and middle plots) on a per-pixel basis, and after regridding in Galactic coordinates (bottom plot).

The average spectral RMS of the final data cube (regridded data converted to Galactic coordinates) is 0.82~K, with a median of 0.78~K and a standard deviation of 0.21~K (at 1~\kms\ resolution).

\begin{figure}
\centering
\includegraphics[width=\hsize, angle=0]{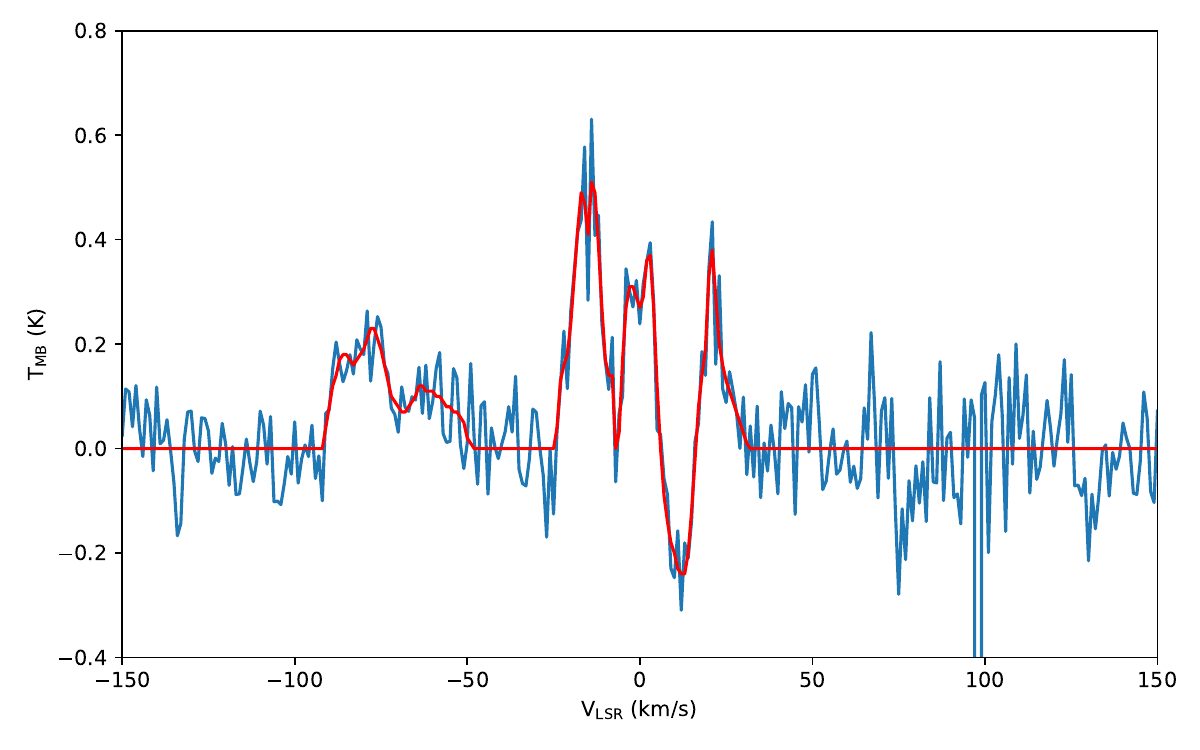}
\caption{Spectrum of the nearby off-position used in the Sgr C [CII] observations (J2000 $17^h44^m14.21^s$, $-28^\deg11'17.5''$). This position, used as the reference for the OTF maps, was observed against a more distant, cleaner position (J2000 $17^h41^m55.21^s$, $-28\deg23'33.3''$) to identify contamination. The spectrum represents the average of all pixels from both polarizations.  The red line represents the spline fitted to the emission features, which was subsequently added back to the Sgr C dataset to correct for the self-subtraction caused by this contamination.}
\label{off}
\end{figure}

To correct for contamination in the nearby off-position, we averaged the spectra for all the pixels in both arrays from a pointed observation of this nearby position against the farther-away reference position. The resulting spectrum is plotted in Fig.~\ref{off}; the RMS residuals with the smooth curve subtracted is 0.08~K. This spectrum shows an apparent absorption at 11.5~\kms, likely due to contamination along the line of sight toward the more distant off-position at this velocity. However, the intensity of this absorption is $\sim -0.25$~K and narrow (5.6~\kms) compared with GC [CII] emission. Therefore, this feature does not affect the bulk of the Sgr~C region (which corresponds to the velocity range $\sim -90$ to $-10$~\kms) and does not affect the results or discussion of this work. We then modeled the off-position spectrum using a spline function only within the velocity ranges that exhibited emission (Fig.~\ref{off}). This modeled off-position spectrum was subsequently added to every spectrum in the database. This approach allowed us to obtain a high signal-to-noise spectrum of the off-position while avoiding the introduction of noise in the final map. However, it introduced uncertainty at the 0.15~K level.

\section{Results}\label{sec:results}

\begin{figure*}
\centering
\includegraphics[width=\hsize, angle=0]{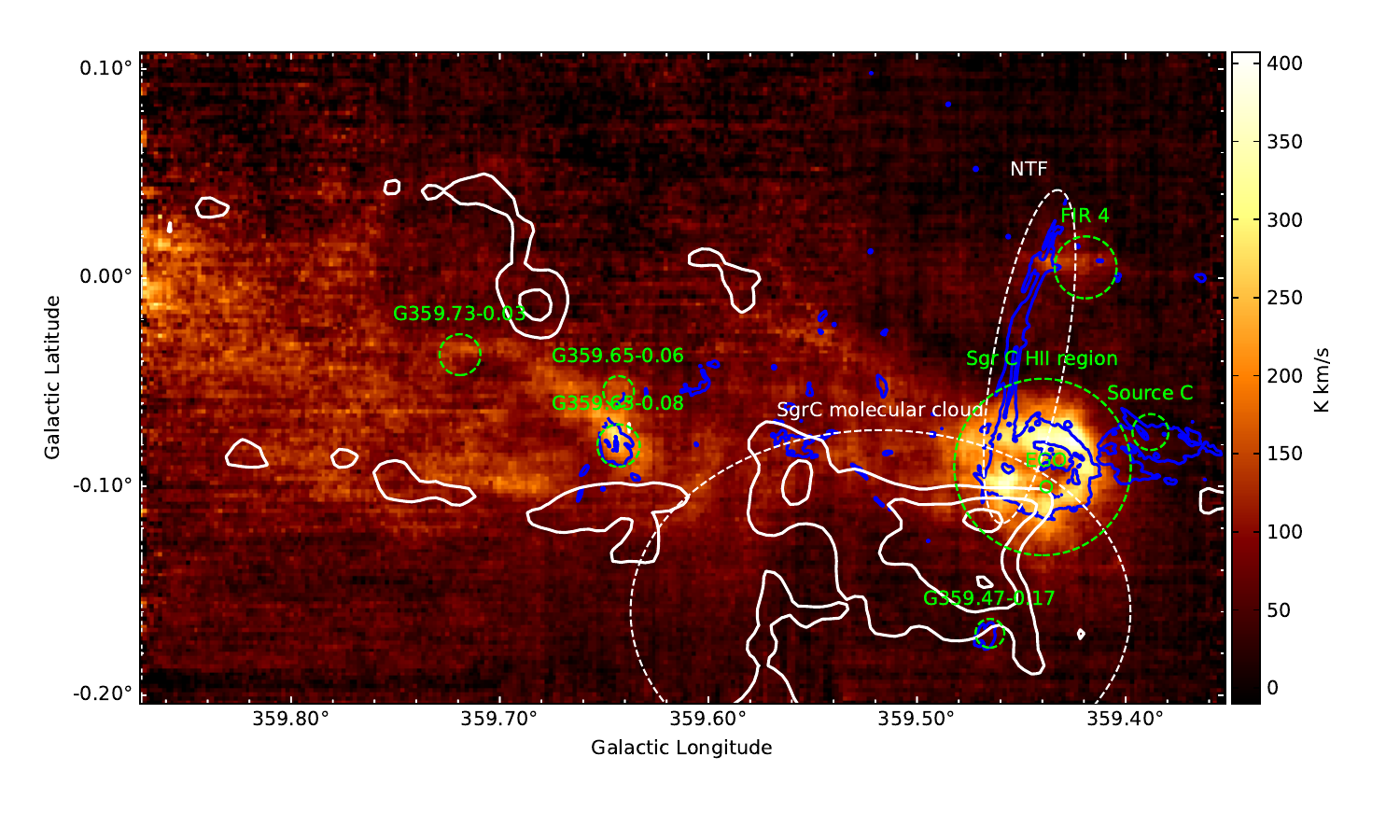}
\caption{[CII] integrated intensity map in the velocity range from
  $-90$ to $-10$ \kms\ in color scale, with previously identified features detected in the [CII] data depicted in green \citep{Liszt_Spiker_1995, Downes_et_al_1980}. In white contours, we show the integrated intensity in the same velocity range of the C$^{18}$O  (2-1) emission (at 40\%, 60\% and 80\% of the peak intensity of 29.21 K \kms) and in blue contours, we show the  MeerKAT radio continuum image centered at 1.28 GHz (23 cm emission) (at 10\%, 50\%, and 90\% of the peak intensity of 0.0134 Jy beam$^{-1}$) from \citet{Heywood_et_al_2022}. We show important features that have no counterpart in the [CII] emission (at the sensitivity level of our observations) in white dashed ellipses. }
\label{identifiedsources}
\end{figure*}

Figure~\ref{identifiedsources} provides an overview of the Sgr~C region. The color image displays the [CII] intensity integrated over the velocity range from $-90$ to $-10$~\kms, with green markings identifying the main sources discussed in this work: the Sgr~C HII region, FIR~4, and Source~C, as well as previously identified sources such as the compact HII regions G359.65--0.08, G359.65--0.06, G359.73--0.03, and G359.47--0.17 \citep{Liszt_Spiker_1995, Downes_et_al_1980}. The Sgr~C HII region is prominent as a circular structure near the western edge, while several bright, compact HII regions identified in previous studies \citep[e.g.,][]{Forster_Caswell_2000, Kendrew_et_al_2013, Lu_et_al_2019a} are visible toward the center of the image. Source~C is not visible in this specific velocity integration as it resides at significantly more blueshifted velocities ($-160$ to $-100$~\kms; see Section 3.1.3). For context, we overlay the C$^{18}$O (2--1) emission (white contours; Riquelme-V\'asquez et al., in prep.) to trace the dense Sgr~C molecular cloud, and the MeerKAT 1.28~GHz radio continuum emission (blue contours; \citealt{Heywood_et_al_2022}) to trace ionized gas.

\begin{figure*} 
\centering
\includegraphics[width=\hsize, angle=0]{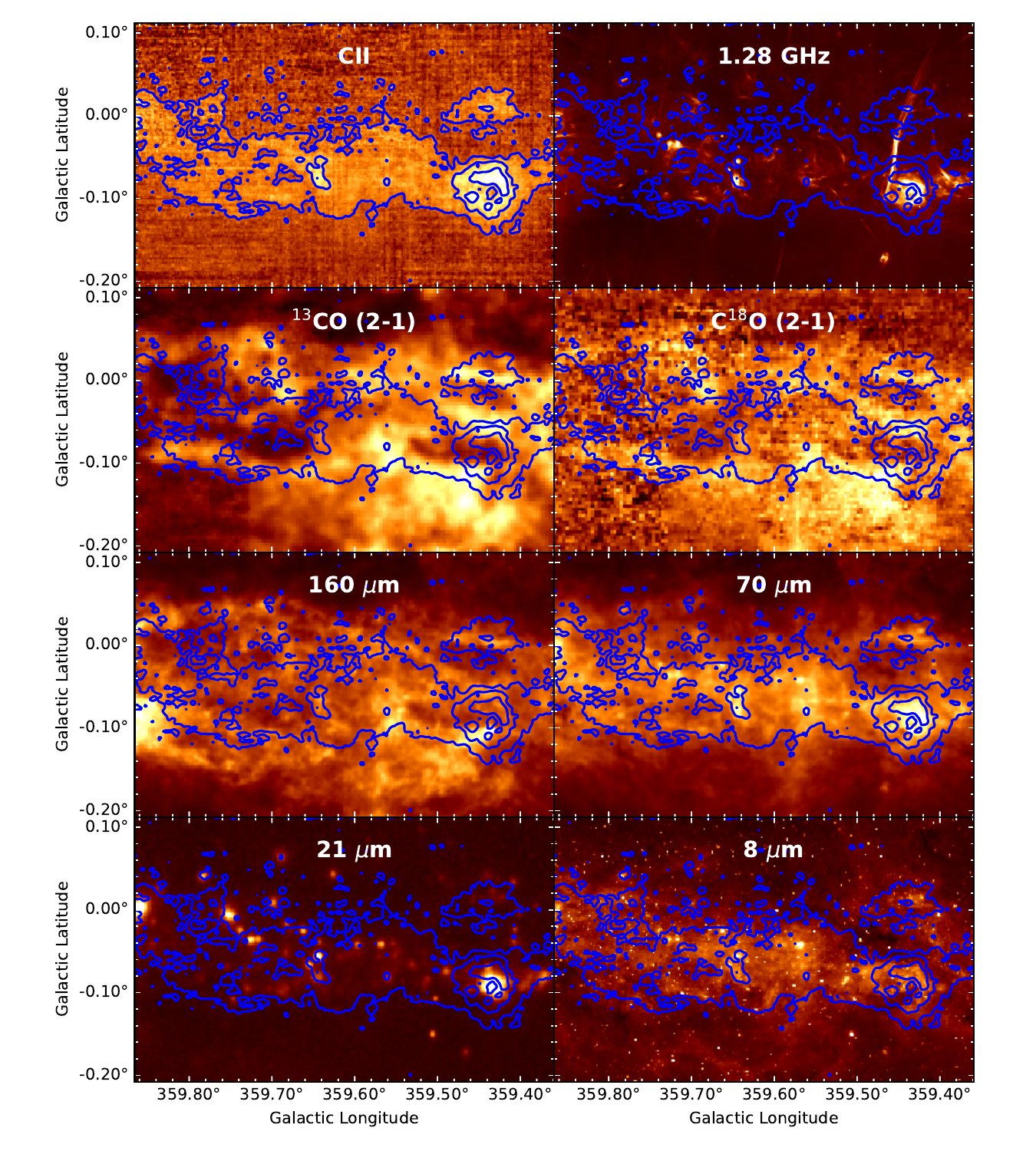}
\caption{Image of the Sgr C region in several tracers. Top left is the
  integrated intensity emission of [CII] in the velocity range from
  $-190$ to $-10$ \kms\ in color scale. Subsequent plots show in color scale the emission from 1.28 GHz (in logarithmic scale) 
\citep{Heywood_et_al_2022}, the integrated intensity of $^{13}$CO (2-1) and C$^{18}$O (2-1) in the same velocity range (Riquelme-V\'asquez et al., in prep.), 160 $\mu$m, 70 $\mu$m \citep{Molinari_et_al_2016}, 21 $\mu$m \citep{Price_et_al_2001}
, and 8 $\mu$m \citep{Stolovy_et_al_2006}
, and in blue contours the [CII] emission at 20\%, 40\%, and 60\% of the intensity peak (495.7 K \kms).}
\label{comparison_-190_-10}
\end{figure*}

Figure~\ref{comparison_-190_-10} places the [CII] emission in the Sgr~C region in spatial context with other tracers. The top-left panel shows the [CII] integrated intensity. The subsequent panels display: the 1.28~GHz radio continuum \citep{Heywood_et_al_2022}; the molecular gas tracers $^{13}$CO (2--1) and C$^{18}$O (2--1) (Riquelme-V\'asquez et al., in prep.); the cold dust emission at 160~$\mu$m and warm dust at 70~$\mu$m from Herschel/PACS \citep{Molinari_et_al_2016}; and compact source tracers at 21~$\mu$m from MSX \citep{Price_et_al_2001} and 8~$\mu$m from Spitzer/GLIMPSE \citep{Stolovy_et_al_2006}.

As in the case for Sgr~B \citep{Harris_et_al_2021}, in the Sgr~C region, there is good overall correspondence between the large-scale distributions of [CII], 1.28~GHz radio continuum, and 70~$\mu$m emission. Conversely, the 160~$\mu$m emission and CO isotopologues exhibit much broader distributions, reflecting the presence of extended, cooler background molecular clouds. 

The bright Sgr~C HII region is prominent in [CII], 1.28~GHz radio continuum, and 70~$\mu$m at the right-center of the images. It lies at the edge of molecular clouds traced in the CO isotopologues, with energy deposition to the cloud from the HII region indicated by a bright spot in the 160~$\mu$m image. As we describe in more detail in Sec.~\ref{HII} and discuss in Sec.~\ref{sec:shellshape}, [CII] and 70~$\mu$m emission toward the Sgr~C HII region have approximately concentric circular shapes with a central depression, with 1.28~GHz emission filling the cavity. A Wolf-Rayet star \citep{Clark_et_al_2021} near the center of the depression likely provides energetic radiation to ionize the HII region, excite a surrounding PDR, and drive expansion. The bright and nearly vertical stripe in the 1.28~GHz image that starts at the edge of the 1.28~GHz emission is one of the brightest nonthermal radio filaments (NTFs) in the Galactic center \citep{Heywood_et_al_2022}. It has no infrared or molecular counterparts. 

The compact bright region close to the center of the [CII] image is the G359.65--0.08 HII region \citep{Liszt_Spiker_1995} with a central velocity of $-41$~\kms\ and an FWHM of 15~\kms.

\begin{figure}
\centering
\includegraphics[width=\hsize, angle=0]{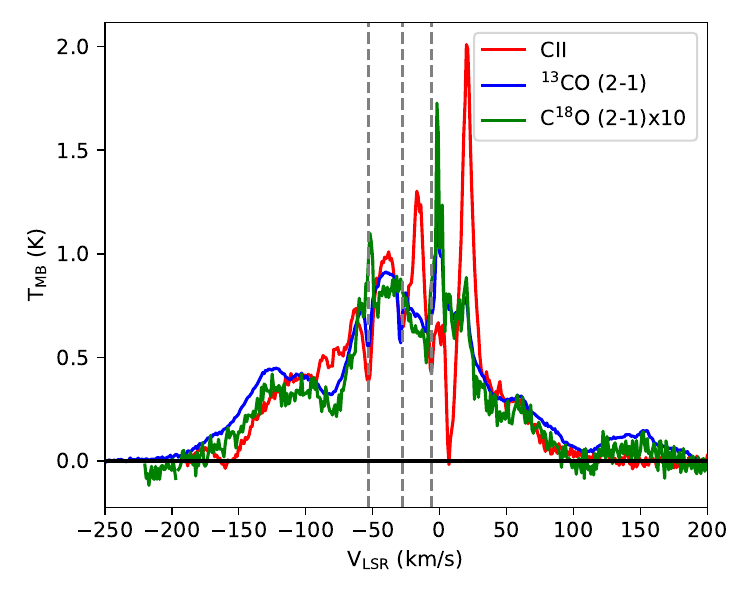}
\caption{Average spectra of the complete Sgr C region for [CII],
  $^{13}$CO (2-1) and C$^{18}$O (2-1). The C$^{18}$O is multiplied by a factor of 10 for easier visualization. Gray dashed lines show the absorptions at velocities $-6$, $-28$, and $-53$ \kms, which correspond to the Crux, Norma, and 3 kpc spiral arms, respectively.}
\label{averagespectra}
\end{figure}

\begin{figure*}
\centering
\includegraphics[width=\hsize, angle=0]{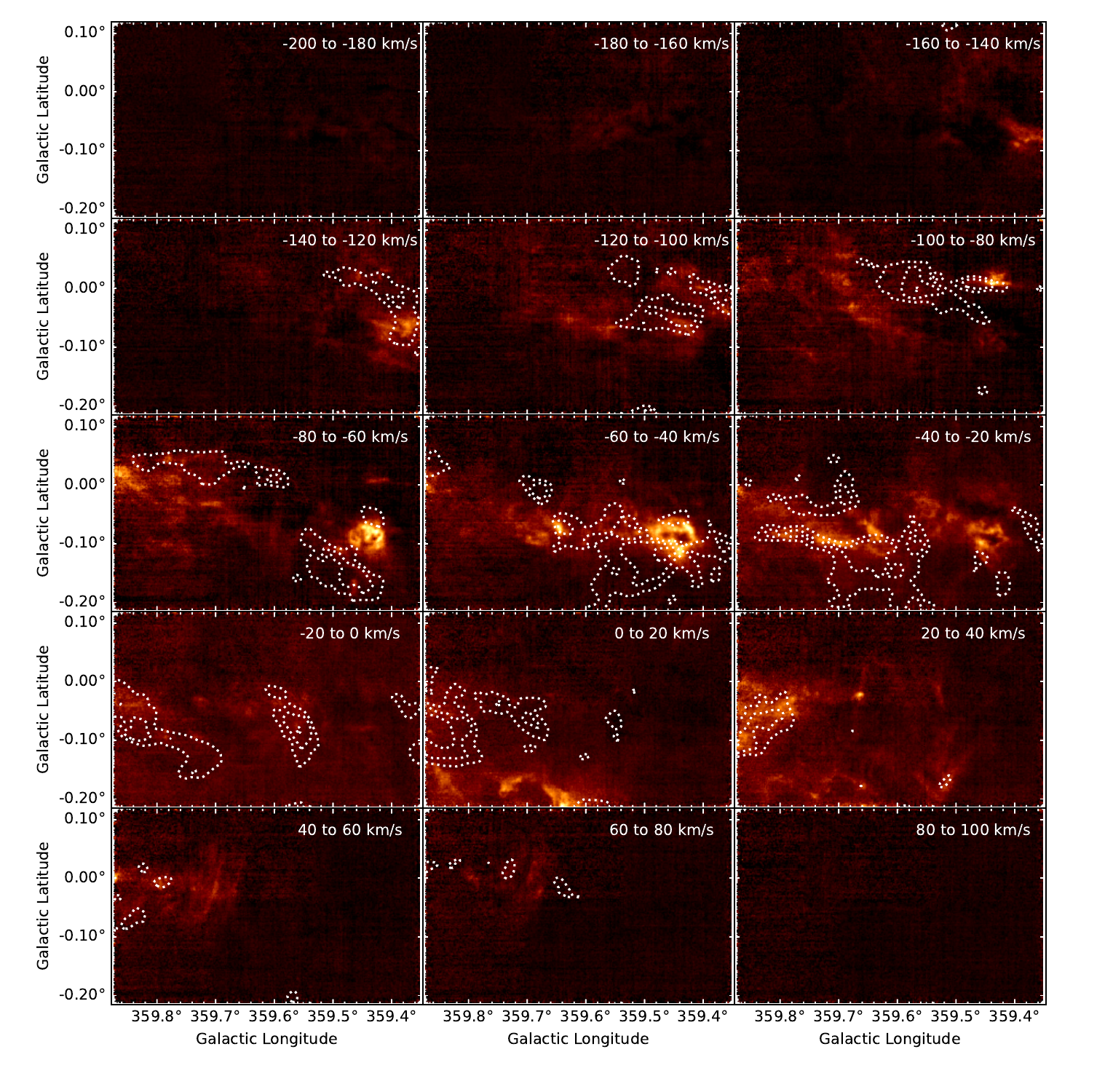}
\vspace{-8mm}
\caption{Channel maps of the velocity-integrated [CII] emission in steps of 20 \kms.  The color scale is linear from $-10$ K \kms\ to 190 K \kms. In white dotted contours, we show the integrated intensity emission in the same velocity range of the $^{13}$CO (2-1) line. We plot the 30\%, 50\%, and 80\% of the intensity peak from the most intense velocity range (108 K \kms\ in the velocity range from $-60$ to $-40$ \kms)}
\label{ciichannelmap}
\end{figure*}

Figure~\ref{averagespectra} compares the [CII] spectrum averaged over the Sgr~C region with profiles of the $J=2$--$1$ rotational transitions of $^{13}$CO and C$^{18}$O from the same field (Riquelme-V\'asquez et al., in prep.). Good overall agreement between the line profiles indicates that [CII] is associated with molecular gas; furthermore, the spatial correspondence between the [CII] emission distribution and the ionizing UV traced by radio continuum (Fig.~\ref{identifiedsources}) places the [CII]-emitting regions at the edges of UV-illuminated molecular clouds. The bulk (63\%) of the [CII] emission arises from gas at negative velocities, following the sense of Galactic rotation (percentages are calculated by comparing the intensity integrated over specific ranges to the total integrated intensity). Specifically, 47\% of the emission lies in the velocity range from $-90$ to $-10$~\kms, and 16\% in the extreme high-velocity range from $-190$ to $-90$~\kms. There is still a substantial amount (12\%) of [CII] emission at forbidden positive velocities between 30 and 130~\kms.
\clearpage
Three [CII] absorption features at velocities of $-6$, $-28$, and $-53$~\kms\ correspond to the Crux, Norma, and 3~kpc spiral arms, respectively. The strong peak at $\sim 20$~\kms\ corresponds to gas in the line of sight, likely unrelated to the GC (see below for further details and Figs.~\ref{ciichannelmap} and \ref{lvdiagram}). We assign the prominent absorption dip at $\sim 10$~\kms\ to [CII] in the diffuse interstellar medium (ISM) along the line of sight. This feature is likely local, given its wide extent of more than $1^\circ$ (it is detected against all background targets, including both line and continuum emission), its narrow linewidth, and its small velocity gradient. While this feature of the diffuse ISM warrants further investigation, its limited velocity range is well separated from the velocities of the Sgr~C complex and does not affect the results of our analysis.

As a complement to the channel maps, Figure~\ref{lvdiagram} presents the [CII] longitude-velocity ($L$--$V$) diagram integrated over the observed latitude range. The overall velocity gradient in the sense of Galactic rotation, apparent as a general tilt toward the lower right in the figure, is partially masked by horizontal dark stripes from the absorption features visible in Fig.~\ref{averagespectra} at $-6$, $-28$, and $-53$~\kms. The overall morphology of the emission resembles the parallelogram observed in CO across the larger Galactic center region \citep{Bitran_et_al_1997, Oka_et_al_1998, Stark_et_al_2004}. This parallelogram feature, seen in both molecular and HI emission, appears at higher velocities \citep[see, e.g., Fig.~10 in][]{Eden_et_al_2020} than the [CII] emission. Here, we identify a diagonal stripe spanning from $-70$ to $-130$~\kms, which likely belongs to the "100~pc stream" \citep{Kruijssen_et_al_2015} rather than the expanding molecular ring (EMR) feature. A bright but fading horizontal stripe at velocities between $+20$ and $+30$~\kms\ arises from galactic emission concentrated at negative longitudes. The Sgr~C HII region appears in the diagram as a bright ellipse at velocities from $-20$ to $-60$~\kms\ at approximately $l = -0.56^\circ$.

\subsection{Prominent [CII] emission components}

\subsubsection{SgrC HII region}\label{HII}
The most intense [CII] feature is associated with the Sgr~C HII region. Figure~\ref{ZoomHII} presents a zoomed view of this region, and Figure~\ref{zoomHIIregion_channel1} in Appendix~\ref{complementaryplots} shows detailed channel maps. As shown in those plots, the emission associated with this feature spans a broad velocity range from $-90$ to $-5$~\kms. Figure~\ref{ZoomHII} displays the [CII] emission in color scale, with contours indicating mid-infrared (MSX 21~$\mu$m) and radio continuum (MeerKAT 1.28~GHz) emission. A panel to the right provides sample spectra toward selected positions. Sgr~C appears as a ring-like structure with a half-intensity radius of approximately $1.8'$. As we describe below, a plausible three-dimensional structure is a thick-walled shell with limb-brightened edges that presents a ring structure in projection.

Figure~\ref{ZoomHII}'s right panel displays spectra toward selected positions around and toward the center of the Sgr~C HII region. The [CII] spectral profiles are broad, extending over a velocity width of 80~\kms, with a line shape that changes rapidly across the source. Varying amounts of absorption in narrow lines at $-53$ and near $-58$~\kms\ distort the line wings but do not affect the peak velocities of the overall emission. The $-53$~\kms\ feature arises from the 3~kpc arm, similar to what is observed in H\,{\sc i} spectra (see Fig.~8 in \citealt{Lang_et_al_2010}). The $-58$~\kms\ absorption has no counterpart in the Galactic disk but matches the central velocity of $^{13}$CO ($2$--$1$) from the Sgr~C molecular cloud, indicating that at least part of that cloud must lie in front of the Sgr~C HII region.

Comparison of the spectra with the distributions of [CII] and radio continuum provides strong evidence that the lineshapes reflect two distinct emission components. Toward all positions, the spectra reveal an intensity component at a relatively constant velocity of $\sim -65$~\kms. This is best illustrated by the spectrum observed toward position 13, the center of the shell (which is also the peak of the HII region); here, the emission peaks at $-65$~\kms, coinciding with the velocity of the HII region traced by the H70$\alpha$ recombination line \citep{Liszt_Spiker_1995}. The lineshapes and linewidths of the two lines are comparable, with 11~\kms\ FWHM in [CII] and 18~\kms\ in H70$\alpha$.

In contrast, the spectrum at position 7, toward low-level thermal radio continuum from the HII region (see also Fig.~\ref{spectral_index}), peaks at a velocity of $-48.2$~\kms, close to the velocity of the surrounding molecular cloud. At these more redshifted velocities, the [CII] emission appears as a clear shell structure in the respective velocity bins (see Fig.~\ref{zoomHIIregion_channel1}). The ring is somewhat patchy, not perfectly symmetric, and appears to break up toward positive longitudes in the direction of the footpoint of the nonthermal filament. Since ionized gas at these velocities is inconspicuous in the H70$\alpha$ recombination line, we conclude that the shell component does not trace the HII region itself, but instead traces [CII] excited in a photodissociation region (PDR) bordering the HII region.

We conclude that the [CII] spectra present a blend of emission from the ionized gas inside the HII region and from a shell-type PDR bordering it. Sgr~C has a structure matching a classic spherical PDR. The source of ionization and the nature of the shell are further discussed in Section~\ref{discussion}.

\begin{figure*}
\centering
\includegraphics[width=\hsize, angle=0]{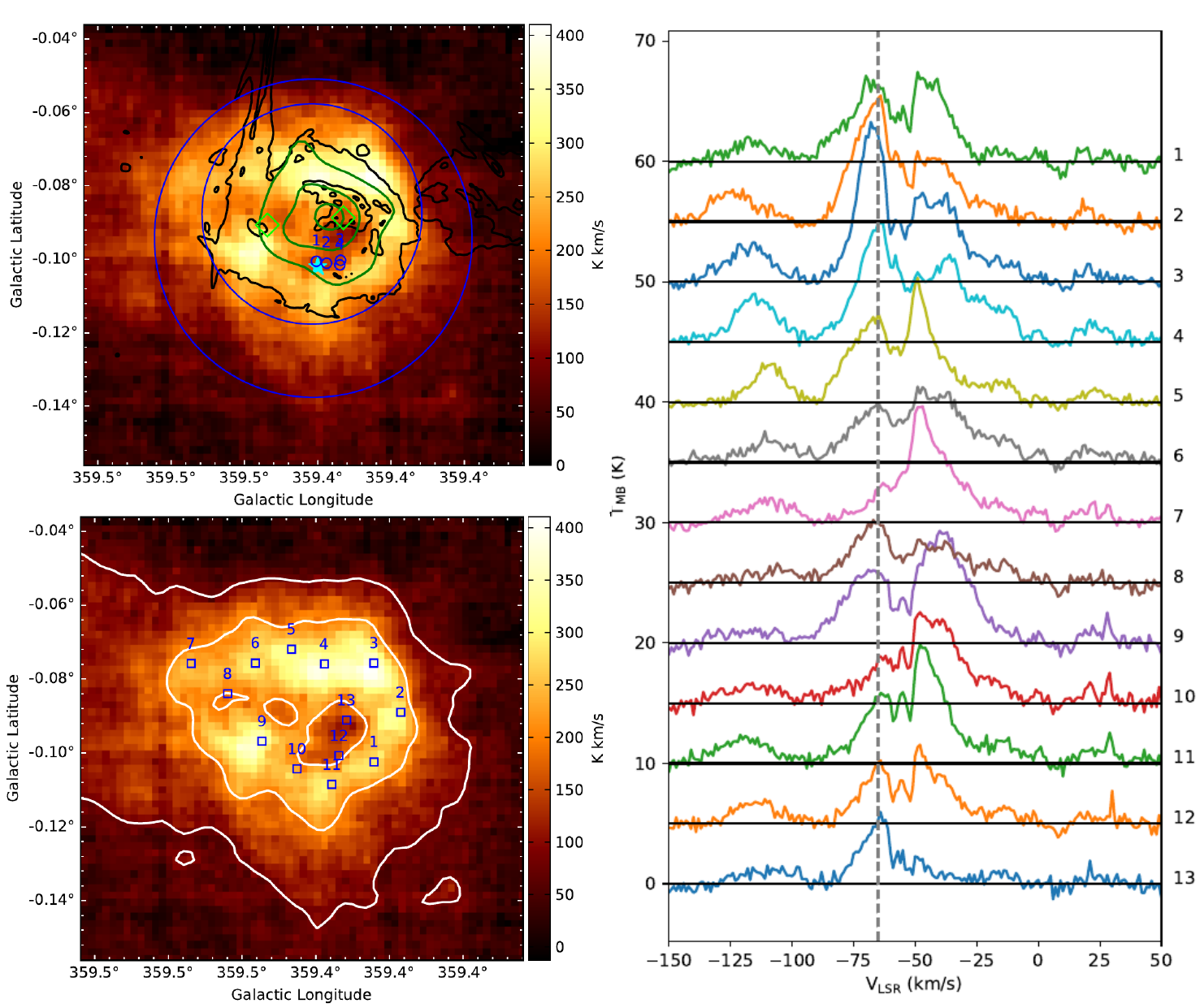}
\caption{Left Top: Integrated intensity emission from $-90$ to $-5$ \kms\ of [CII] emission towards the HII region. The green contours show 20\%, 50\%, and 80\% of the MSX 21 $\mu$m emission intensity peak \citep{Price_et_al_2001}. In black contours, we show the 10\%, 50\%, and 80\% of the 1.28 GHz emission intensity peak
\citep{Heywood_et_al_2022}. The big blue circle of radius 2.6 arcmin shows the region considered to compute the [CII] mass, which corresponds to 20\% of the intensity peak. The smaller blue circle of radius 1.8 arcmin shows 50\% of the intensity peak. In tiny blue circles, we show the HII regions identified by \citet{Lu_et_al_2019a, Forster_Caswell_2000} following the notation of \citet{Lu_et_al_2019a}, where 1, 2, 3, and 4 refer to H1, H2, H3, and H4 from their Table 5. Regions 3 and 4 were studied by \citet{Kendrew_et_al_2013}. The cyan star corresponds to the protocluster G$359.44-0.102$ \citep{Crowe_et_al_2025}. The light-green diamonds show the position of the Wolf-Rayet stars \citep{Clark_et_al_2021}.
Left Bottom: Same as the top figure, showing in white contours the 20\% and 50\% contours of the intensity peak of the [CII] emission and in blue squares the positions selected to extract the [CII] spectra. 
Right: Selected [CII] spectra over the beam size in the HII region. Gray dashed vertical lines mark the velocity $-65$ \kms\ as a reference, which corresponds to the velocity of the peak of the H70$\alpha$ recombination line.}
\label{ZoomHII}
\end{figure*}

 \subsubsection{FIR 4}
This source was previously identified as an HII region by \citet{Odenwald_Fazio_1984} and described further by \citet{Liszt_Spiker_1995}. Figure~\ref{FIR4} shows the [CII] intensity map integrated from $-90$ to $-85$~\kms\ as blue contours over the 1.28~GHz radio continuum emission in color, along with the spectrum integrated over the source. The spectrum indicates a velocity of $-86$~\kms\ toward the center of the source, with an FWHM linewidth of 16~\kms. This velocity and the H\,{\sc i} absorption against the continuum \citep{Lang_et_al_2010} place FIR~4 in the Galactic Center.

As \citet{Liszt_Spiker_1995} noted, the nonthermal filament brightens as it crosses the boundary of FIR~4. \citet{Zhao_et_al_2025} propose that FIR~4 is interacting with the nonthermal filament, a scenario supported by MeerKAT observations. A detailed analysis of the [CII] emission, integrated between $-90$ and $-85$~\kms\ and overplotted on the 1.28~GHz radio continuum, shows a distinct intensity increase toward the east--west thermal ridge (see Fig.~\ref{FIR4}, left). Unfortunately, in contrast to our analysis of the Sgr~C shell (Sect.~\ref{discussion}), our data lack the angular resolution needed to search for kinematic evidence of a physical interaction between the FIR~4 HII region and the nonthermal filament.

To estimate the amount of gas associated with the [CII] emission from FIR~4, we consider the area enclosed by the 20\% contour of the intensity peak (197.6~K~\kms) from the intensity map integrated over the velocity range from $-100$ to $-62$~\kms, which corresponds to the complete velocity range of the [CII] emission in this feature (see Fig.~\ref{FIR4}, right). Using the same methodology as in Appendix~\ref{sec:columndensity} and Discussion Sect.~\ref{sec:shell-kinematics}, we obtain a column density of $N(\text{C}^+) = 2.74(\pm 0.78) \times 10^{18}$~cm$^{-2}$ and a C$^+$ mass of $6.3(\pm 1.8)~M_{\odot}$. These values imply lower limits for the hydrogen column density and mass of $N(\text{H}) = 9.15(\pm 2.62) \times 10^{21}$~cm$^{-2}$ and $M(\text{H}) = 1.75(\pm 0.5) \times 10^{3}~M_{\odot}$.

\begin{figure}
\centering
\includegraphics[width=\hsize]{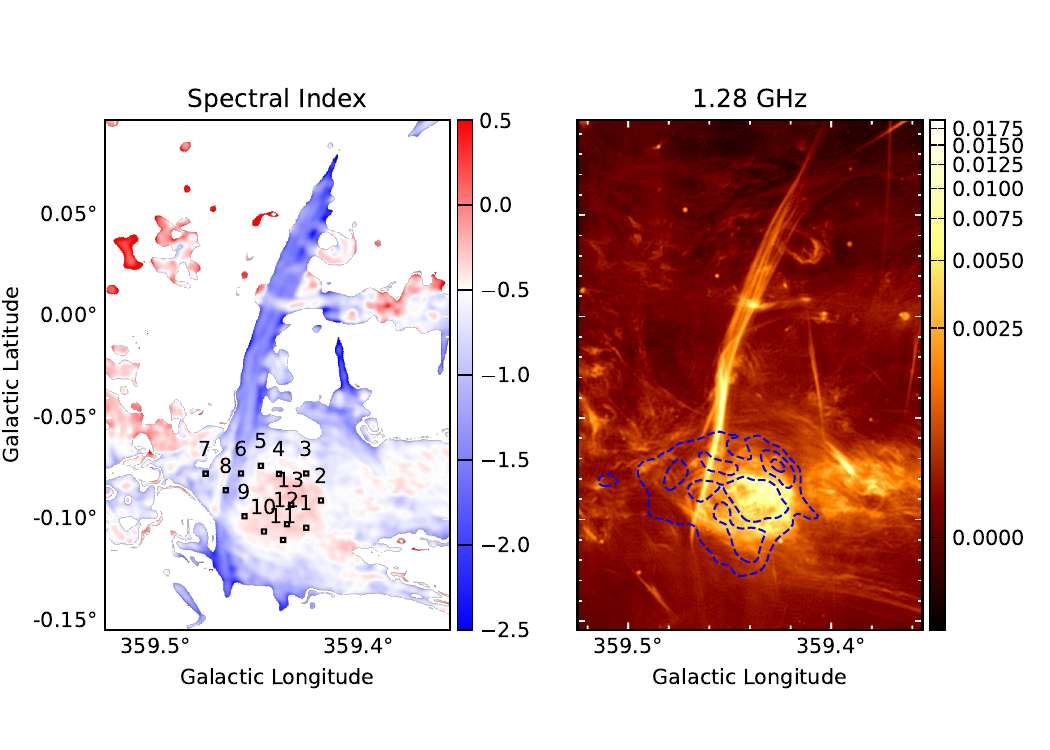}
\caption{Left: Plot of the spectral index with the positions selected in Fig. \ref{ZoomHII} overplotted. Right: 1.28 GHz MeerKAT emission over the same region with the [CII] shell contours overplotted.}
\label{spectral_index}
\end{figure}

\begin{figure*}[t]
\centering
\hbox{
\includegraphics[width=0.5\hsize, angle=0]{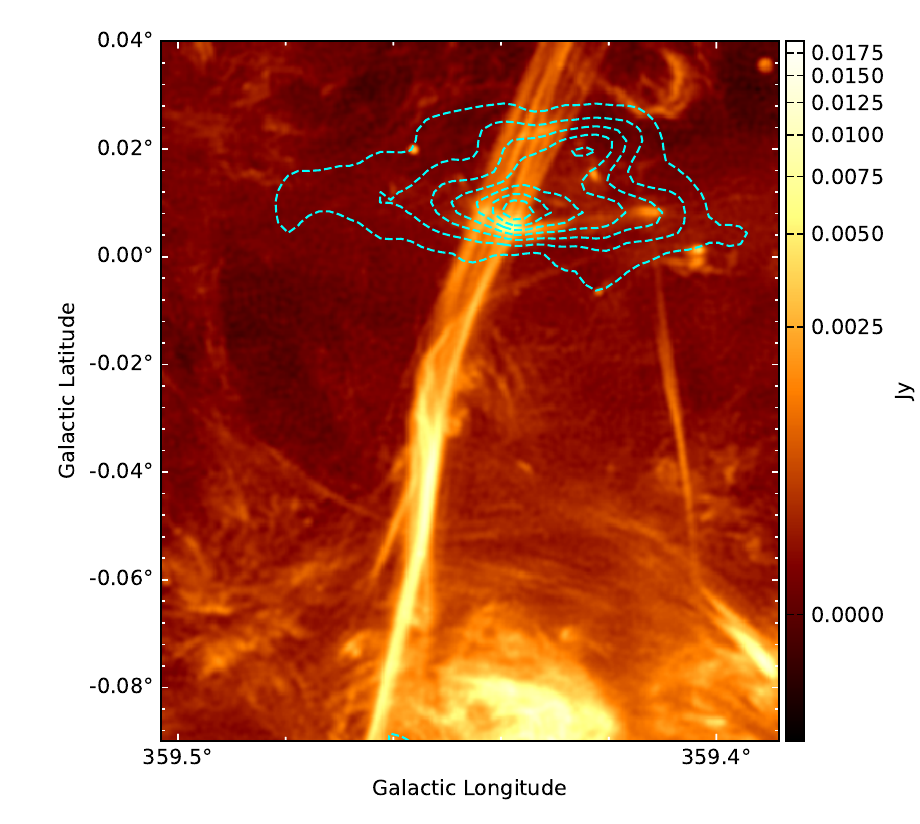}
\includegraphics[width=0.5\hsize, angle=0]{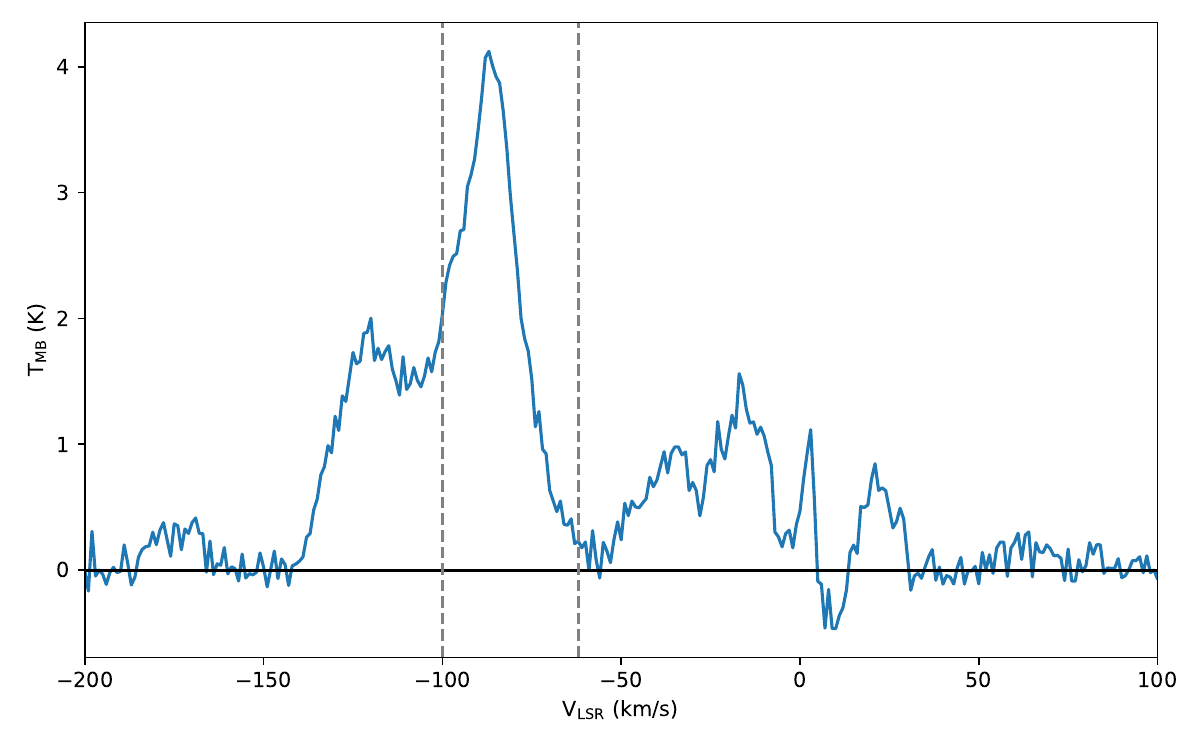}
}
\caption{Left: MeerKAT 1.28 GHz continuum emission \citep{Heywood_et_al_2022} of the nonthermal filament and the FIR 4 region. Cyan contours represent the [CII] integrated intensity over the velocity range $-90$ to $-85$ \kms. Contour levels start at 10\% and increase to 90\% of the peak intensity (72 K \kms) in steps of 10\%. Right: [CII] average spectrum over the FIR4 region.  Gray dashed vertical lines mark the velocities $-100$ and $-62$ \kms\ as a reference. }
\label{FIR4}
\end{figure*}

\subsubsection{Source C}

\begin{figure}
 \centering
\includegraphics[width=\hsize, angle=0]{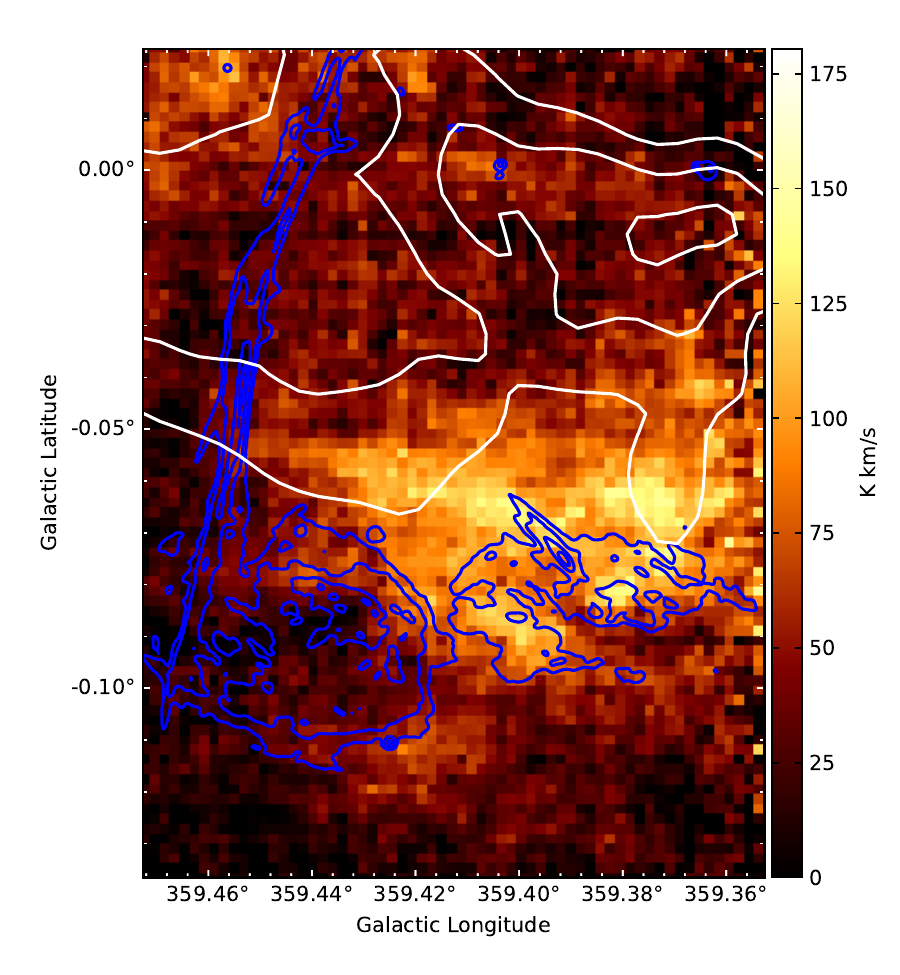}
\caption{Zoom towards Source C in [CII]. Integrated intensity in the velocity range $-160$ to $-100$ \kms. In blue contours, we show 10\%, 20\%, 50\%, and 80\% of the 1.28 GHz emission intensity peak \citep{Heywood_et_al_2022}. In white contours we show 50\%, 70\%, and 90\% of the $^{13}$CO integrated intensity peak (131 K \kms) in the same velocity range. }
\label{sourceC}
\end{figure}

The "Source~C" HII region (Fig.~\ref{sourceC}) was identified as a distinct HII region situated adjacent to the primary Sgr~C HII region and the Sgr~C nonthermal filament by \citet{Lang_et_al_2010}. Its nature was further elucidated through its H\,{\sc i} absorption spectrum, which bears a strong resemblance to other sources within the Sgr~C complex and prominently features absorption lines near LSR velocities of $-60$, $-100$, and $-126$~\kms. Based on the analysis of atomic, molecular, and ionized gas velocities, \citet{Lang_et_al_2010} suggested that Source~C is likely positioned behind atomic gas associated with a molecular cloud at $-65$~\kms; this intervening atomic gas is also thought to be physically connected to the Sgr~C NTF. Similarly, atomic gas linked to another molecular cloud at $-100$~\kms\ is probably located on the near side of Source~C and may share a physical relationship with both Source~C and the Sgr~C NTF. \citet{Liszt_Spiker_1995} found that it is stippled with thermal radio continuum emission, as can be seen in Fig.~\ref{spectral_index}. This source appears in [CII] emission as a diffuse patch in the velocity range from $-160$ to $-100$~\kms, significantly blueshifted relative to the velocities of the Sgr~C HII region ($-65$ to $-45$~\kms; see Fig.~\ref{ciichannelmap}). Although molecular emission traced by CO and its isotopologues is widespread throughout the Sgr~C region, Fig.~\ref{sourceC} shows that CO makes a negligible contribution within this velocity range and has no obvious morphological correspondence with the [CII] emission.

\subsection{Other C$^+$ and molecular emission regions}
\begin{figure}
\centering
\includegraphics[width=\hsize, angle=0]{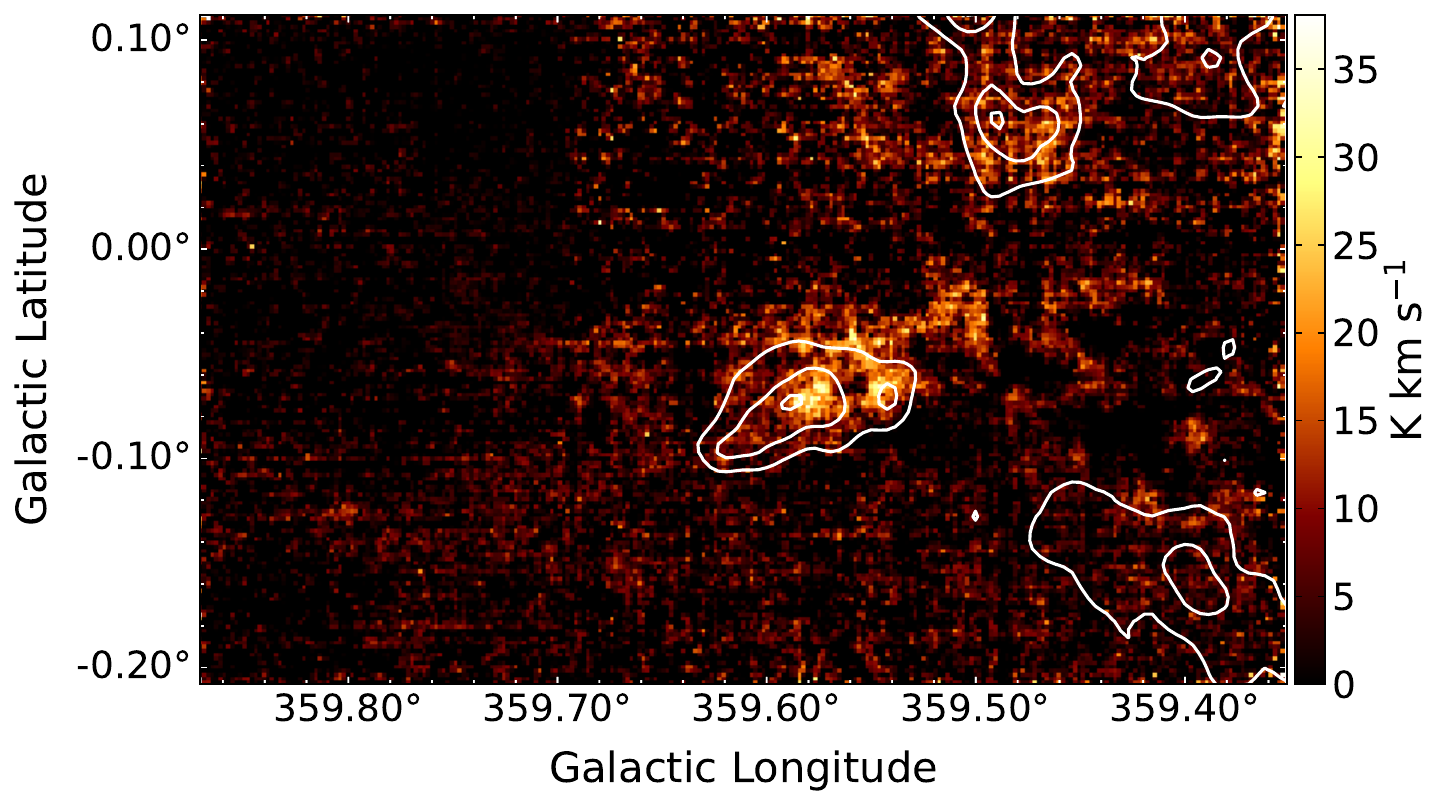}
\caption{Integrated intensity of the [CII] emission in the velocity range from $-190$ to $-160$ \kms, and in white contours the 20\%, 50\% and 80\% of the $^{13}$CO $(2-1)$ integrated intensity peak (39.97 K \kms) for each velocity range.}
\label{hvc}
\end{figure}

Faint [CII] emission with a strong $^{13}$CO ($2$--$1$) molecular counterpart is visible toward a high-velocity cloud ($-190$ to $-160$~\kms) in Fig.~\ref{hvc}. At this Galactic longitude, the velocity range corresponds to emission from the ``CO parallelogram'' that has been attributed to gas flows along orbits appropriate for the Galactic center's bar potential \citep[e.g.,][]{Sormani_et_al_2018, Sormani_et_al_2019}. We also expect the existence of high-velocity clouds on the positive-velocity segments of their orbits. We do not detect [CII] emission from these at the sensitivity level of our observations, and the source AFGL~5376 \citep{Uchida_et_al_1994} is not covered by our survey.

\section{Discussion}\label{discussion}
In view of the volume and detail of the data presented, in this work we focus on Sgr~C, the most prominent source in this part of the survey. With the release of the survey data, the interested community is invited to participate in further analysis of these observations. 

\subsection{Origin of the [CII] emission toward the Sgr~C HII region}\label{sec:shellshape}

As highlighted in previous sections, the most remarkable [CII] feature in this part of our survey is the emission associated with the extended Sgr~C HII region. In addition to [CII] emission originating from within the HII region itself, we find a striking shell-type feature: a PDR bordering the HII region, the velocity field of which is consistent with that expected for the kinematics of an expanding (likely incomplete) shell. In the following discussion, we will first constrain the kinematics and energetics of the feature and then discuss possible driving mechanisms.

\subsubsection{Expanding shell kinematics}
\label{sec:shell-kinematics}

To investigate the hypothesis that the [CII] emission is tracing an expanding shell, we quantitatively compared the observed kinematics with simple models of spherical expansion. We first estimated the spatial dimensions of the shell by analyzing the velocities at which the observed ring structure is largest and most intense. In an expanding spherical shell, these velocities correspond to the shell's systemic velocity, where the expansion is roughly perpendicular to the line of sight in a radial direction from the center of the ring; the corresponding spatial structure represents a 2D projection of the shell. We constructed a [CII] velocity-integrated map of the Sgr~C HII region from $-50$ to $-45$~\kms\ (a range that avoids contamination from the 3~kpc arm), which is shown in Fig.~\ref{fig:pv_shell} (\emph{left}). 

\begin{figure*}
\centering
\includegraphics[width=\hsize, angle=0]{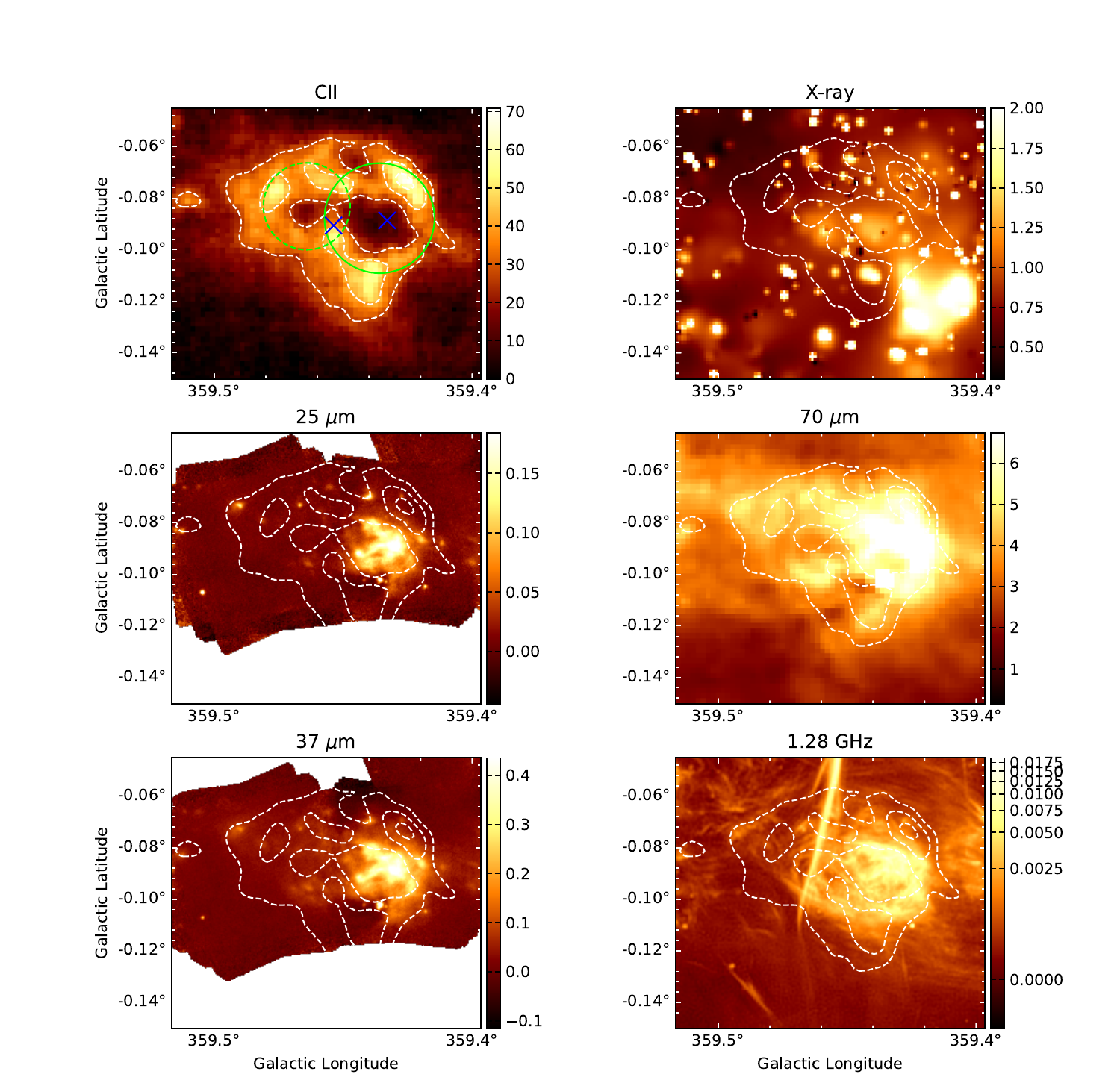}
\caption{Shell/bubble features in different tracers. The top left panel shows the [CII] emission integrated over the velocity range from $-50$ to $-45$ \kms. The green continuous circle shows the proposed expanding shell, which has a clear counterpart in all species shown in the other images. The green dashed circle shows the bubble feature, which does not show a counterpart in the ionization tracers. The blue crosses show the location of the two known WR stars \citep{Clark_et_al_2021}.
The subsequent panels show the 1 - 4 keV X-ray mission \citep{Wang_2021}, FORCAST 25 $\mu$m continuum \citep{DeBuizer_et_al_2025, FORCAST_data}, FORCAST 37 $\mu$m continuum \citep{DeBuizer_et_al_2025, FORCAST_data}, and the MeerKAT 1.28 GHz \citep{Heywood_et_al_2022} in logarithmic scale, with the [CII] emission overplotted with white contours at 40\%, 60\% and 80\% of the intensity peak (70 K \kms).}
\label{shell_tracers}
\end{figure*}

\begin{figure*}
\centering
\includegraphics[width=1.0\hsize]{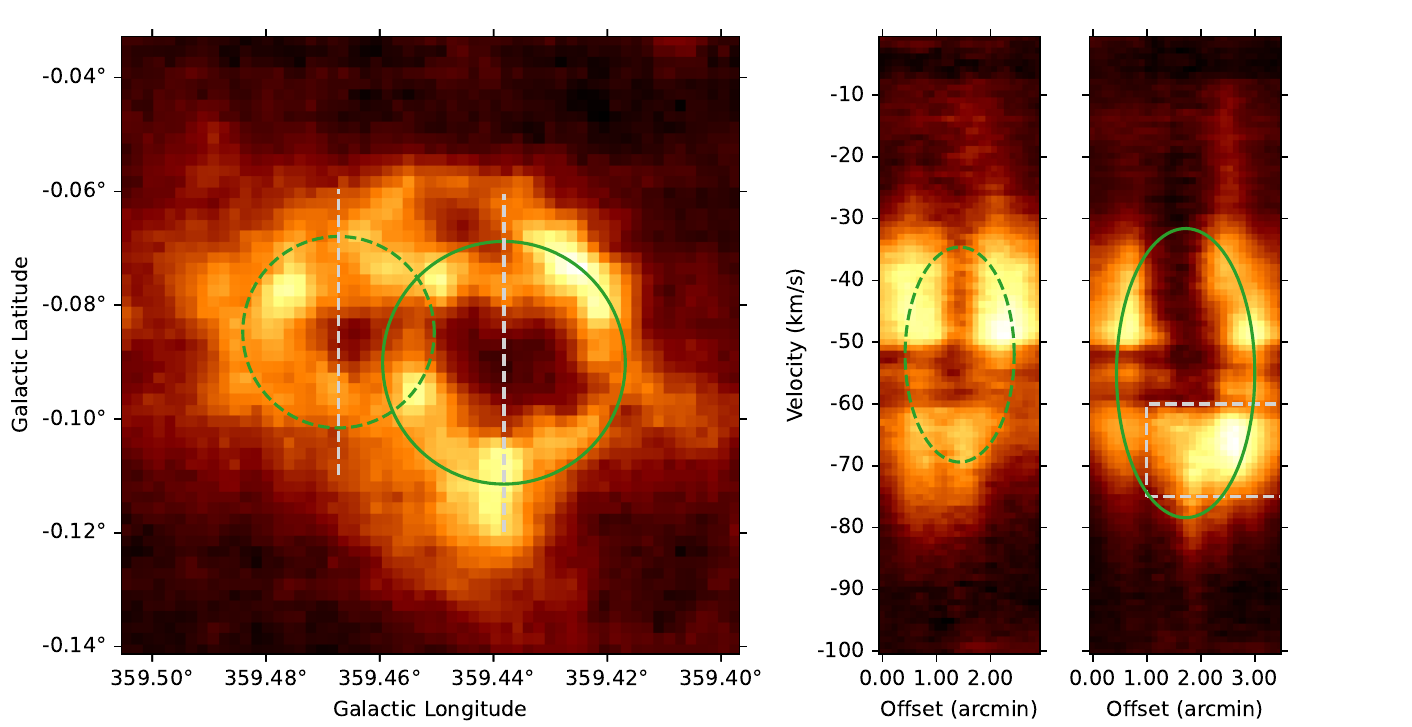}
\caption{Spatial and kinematic characterization of the [CII] emission of the Sgr~C HII region as an expanding shell. \emph{Left:} [CII]  velocity-integrated map from $-50$ \kms\ to $-45$ \kms. The green circle represents a weighted ODR fit to all pixel coordinates with emission larger than $30\%$ of the peak intensity. The dashed vertical line crossing the circle center indicates a reference path along which one p-v diagram was extracted for the shell. The final p-v diagram was obtained by averaging 9~different p-v diagrams using paths with a width of $30''$ and angles between $\pm 45\deg$ with respect to the vertical dashed line. \emph{Middle:} The final averaged p-v diagram for the eastern bubble. \emph{Right:} The final averaged p-v diagram for the shell. The green ellipse shows a weighted ODR fit in p-v space. The dashed rectangle indicates the velocity range and offsets where the emission is dominated by the HII region, which was excluded in the fitting.
}
\label{fig:pv_shell}
\end{figure*}

While the morphology and kinematics of the channel maps (Figs.~\ref{ciichannelmap} and \ref{zoomHIIregion_channel1}) reveal two or more voids of low-level [CII] emission, only one hosts strong radio and mid- to far-infrared emission, implying local sources of massive energy release (namely Sgr~C; see Fig.~\ref{shell_tracers}). This is also evident in the FORCAST/SOFIA images \citep{FORCAST_data, Zhao_Sub_2025}. The fractional drop in intensity---defined as the dimensionless ratio between the intensity at the center of the left void (the bubble) and the intensity at the center of the right void (the shell)---is 0.11 at 25~$\mu$m and 0.15 at 37~$\mu$m in Fig.~\ref{shell_tracers}. In the following, we refer to the right void as the ``shell'' (solid circle in Fig.~\ref{shell_tracers}), while the other void is referred to as the ``bubble'' (dashed circle). 

In the following sections, we analyze the kinematics of this prominent shell feature to test the hypothesis of an expanding shell of gas driven by feedback from the Sgr~C cluster of massive stars.

To characterize the expanding shell, we assumed that the projection of an optically thin, expanding 3D shell is a ring, and that the projected ring on the plane of the sky has a circular shape. We performed a weighted orthogonal distance regression (ODR) to fit the size of the shell, as well as the size of the bubble, shown with a dashed circle in Fig.~\ref{fig:pv_shell}. The input data were simply the $x_i, y_i$ pixel coordinates of the map for all pixels with emission $T_i$ larger than 30\% of the peak intensity. This threshold was chosen to reliably trace the ridge of the shell structure while excluding the surrounding diffuse background and noise. We used the emission values $T_i$ as weights in the minimization to provide more relevance to the pixels with higher intensity. To ensure convergence, each fit was performed independently in a rectangular cut-out enclosing the corresponding ring. The resulting circles have radii of $1.28'$ for the shell (shown as a continuous circle) and $1.01'$ for the eastern bubble, as shown in Fig.~\ref{fig:pv_shell} (\emph{left}).

We then proceeded to derive the velocity information of the tentative expanding shell using the fitted spatial structures and conveniently extracted position--velocity ($p$--$v$) diagrams. Geometrically, for an expanding spherical surface with radius $R$ and expansion velocity $V_{\text{exp}}$, a $p$--$v$ diagram across a projected diameter of the sphere results in an ellipse with semi-axes $R$ and $V_{\text{exp}}$. Since the projected ring is circular, the symmetry allowed us to choose nine different angles between $\pm 45^\circ$ with respect to Galactic north for the diametrical line for $p$--$v$ extraction. We did not use larger angles in order to minimize confusion between the shell and the bubble feature adjacent to it. In Figure~\ref{fig:pv_shell}, the resulting averaged $p$--$v$ diagram for the shell is shown in the \emph{right} panel, with the corresponding best fit. The best-fit parameters are $V_{\text{sys}} = -55.0 \pm 1.6$~\kms\ and $V_{\text{exp}} = 23.4 \pm 2.8$~\kms.

In the fitting, we explicitly excluded the velocity range corresponding to the HII region ($-75$ to $-60$~\kms) in the offsets where it is dominant (shown as a rectangle in Fig.~\ref{fig:pv_shell}) to avoid contamination from the ionized gas component. While absorption features are present in the spectra (e.g., at $-53$~\kms), the ODR fitting to the emission ridge is robust against these narrower absorption dips. We found the fitted parameters to be sensitive to the choice of the excluded data range. Therefore, we conservatively estimated the uncertainties given above by repeating the fit twice: once using all the data, and once with a larger zone of exclusion (velocities from $-75$ to $-50$~\kms\ for all offsets, in order to also exclude the velocity range contaminated by spiral arms). For comparison, purely statistical uncertainties derived from bootstrapping are much smaller, on the order of 0.5~\kms.

We have also investigated the $p$--$v$ diagram of the bubble feature. In the hypothetical case that this feature represents an expanding shell, the best-fit parameters are $V_{\text{sys}} = -52.0 \pm 0.3$~\kms\ and $V_{\text{exp}} = 17.4 \pm 0.4$~\kms\ (uncertainties estimated by bootstrapping only). However, as explained previously, the lack of evidence for any major associated driving source indicates that this "bubble" might not necessarily represent an internally driven expanding shell (nevertheless, the kinematics may show signpost features of expansion; see Fig.~\ref{zoomHIIregion_channel1}).

With a radius of $1.28'$, which corresponds to a diameter of 6.1~pc at a Galactocentric distance of 8.2~kpc \citep{Gravity_Collaboration_2019}, and an expansion velocity of 23.4~\kms, we derive a dynamical age for the shell of only 0.13~Myr. This dynamical age estimate assumes a constant expansion velocity, but it represents a lower limit if the shell has been continuously accelerated by strong stellar winds (see Sect.~4.2.2). Conversely, it serves as an upper limit if a supernova blast wave initiated the shell expansion and the shell has been slowing down as it snowplows the dense material in the surrounding ambient medium as a result of momentum conservation. 

To estimate the amount of gas associated with the [CII] emission, we follow the formulations by \citet{Goldsmith_et_al_2012} and \citet{Crawford_et_al_1985}, adapted to the region's expected physical conditions (see Eqs.~A.1--A.3 in Appendix~A for details). Several assumptions must be made. First, we make the reasonable assumption that the excitation of the [CII] emission associated with the shell is PDR-type (meaning: caused by collisions with molecular hydrogen). Second, because of its blend with emission from the HII region and contamination due to line-of-sight absorption, we assume symmetry in velocity space. We assume that the intensity integrated over the redshifted gas only (from $-50$ to $-5$~\kms) represents half of the total. 

Using Eq.~A.3, we calculate an average column density of $N(\text{C}^+) = 3.93(\pm 1.12) \times 10^{18}$~cm$^{-2}$ and a C$^+$ mass of $24.6(\pm 7.0)~M_{\odot}$. These values are computed within a radius of 1.5 times the radius of the shell, i.e., $1.92'$ (4.6~pc); this integration radius was chosen to encompass the extended diffuse emission associated with the PDR that lies beyond the limb-brightened ridge. Assuming next that all carbon is in the form of C$^+$ in this layer and that the fractional abundance is $[\text{C}]/[\text{H}] \leq 3 \times 10^{-4}$ \citep{Genzel_et_al_1990}, we derive a lower limit for the total hydrogen column density of $N(\text{H}) = 1.31(\pm 0.37) \times 10^{22}$~cm$^{-2}$ and for the hydrogen mass of $6.85(\pm 1.96) \times 10^3~M_{\odot}$. Multiplying the derived value by 2 to account for the blueshifted side, we obtain a total carbon shell mass of $49.2(\pm 14.2)~M_{\odot}$ and a hydrogen shell mass of $13.7(\pm 3.92) \times 10^3~M_{\odot}$. Scaling these derived values by a factor of 1.38 to account for elements heavier than hydrogen \citep{Asplund_et_al_2009}, the total gas mass is $1.89(\pm 0.54) \times 10^4~M_{\odot}$. 

With an expansion velocity of 23.4~\kms, we derive a kinetic energy ($\frac{1}{2} M_{\text{H}} v_{\text{exp}}^2$) for the expanding shell of $1.03(\pm 0.38) \times 10^{50}$~erg and a momentum ($M_{\text{H}} v_{\text{exp}}$) of $4.4(\pm 1.4) \times 10^5~M_{\odot}$~\kms. 

As discussed in Appendix~\ref{sec:columndensity}, the [CII] excitation and the derived mass depend on the local physical conditions and line opacity. A very conservative (factor of 5) lower limit on the mass, kinetic energy, and momentum is derived by assuming optically thin emission under thermalized excitation conditions (Eq.~A.4).

In comparison, the mass associated with the ionized gas in the HII region is an order of magnitude smaller than the initial estimation. Integrating over the velocity range from $-70$ to $-60$~\kms\ and following Eq.~A.6, we calculate $N(\text{C}^+) = 2.67(\pm 0.06) \times 10^{17}$~cm$^{-2}$ and a $C^+$ mass of $1.68(\pm 0.04)~M_{\odot}$. With the same assumptions as above, $N(\text{H}) = 8.9(\pm 0.16) \times 10^{20}$~cm$^{-2}$ and the hydrogen mass $M_{\text{H}} = 466.9(\pm 8.3)~M_{\odot}$. We emphasize that due to potential self-absorption and foreground absorption, these values should be considered conservative lower limits.

\begin{table}
\caption{Properties of the Sgr C shell}             
\label{shell_properties}      
\begin{center}                         
\begin{tabular}{l l }        
\hline
shell radius (pc)             &  $\sim$ 3.0  \\ 
velocity of expansion (\kms)  & 23.4        \\ 
dynamical age (Myr)           & 0.13        \\ 
mass (M$_{\odot}$)            & $1.89 (\pm0.54) \times 10^4$  \\
kinetic energy (ergs)         & $1.03 (\pm 0.38)\times 10^{50}$ \\
momentum (M$_{\odot}$ \kms)   & $4.4 (\pm 1.4) \times 10^5$     \\
\hline
\end{tabular}
\end{center}
\small{The gas mass is derived for subthermally excited, moderately optically thick [CII] emission. In the optically thin, thermalized limit, mass, kinetic energy, and momentum would result in a factor of 5 lower. See text for details.}
\end{table}

\subsection{Physical mechanisms driving the expanding shell}
To investigate the origin of the [CII] shell caused by stellar feedback, we calculated an uncorrected Lyman continuum photon production rate, $N'_{{\rm LyC}}$, following \citet{Mezger_et_al_1974}. We used a radio flux density of 6.8~Jy (at 4.8~GHz) from \citet{Law_et_al_2008} and an electron temperature of $6800 \pm 700$~K \citep{Wink_et_al_1983}\footnote{There are several values for $T_{\rm e}$ in the literature. We prefer the value from Wink et al. because their observations at 14.7~GHz, performed with a small $1'$ beam, are likely best coupled to the thermal emission of the H\,{\sc ii} region.}. The derived $N'_{{\rm LyC}}$ is $5.3 \times 10^{49}$~s$^{-1}$. This value should be considered an upper limit because the 6.8~Jy flux represents the total thermal emission from Sgr~C, which has an effective radius of $2.8'$ \citep[Table~5 in][]{Law_et_al_2008}. This is significantly larger than the $1.28'$ radius of the [CII] shell derived in our analysis. 

This classifies Sgr~C as one of the brightest (classified as "giant") H\,{\sc ii} regions in our Galaxy. \citet{Liszt_Spiker_1995} argue that the H\,{\sc ii} region may host the equivalent of an O5.5~V star; based on the bolometric luminosity of $L = 1.1 \times 10^6~L_{\odot}$ \citep{Odenwald_Fazio_1984}, \citet{Lang_et_al_2010} proposed an O4~V star.

As discussed by \citet{deBuizer_et_al_2022}, the brighter an H\,{\sc ii} region is, the more likely it is to be powered by a cluster of young massive stars. Indeed, according to non-LTE line-blanketed atmospheric models for O-type stars \citep{Martins_et_al_2005}, the Lyman continuum flux of a single main-sequence O4~V star, $\log(N_{{\rm LyC}}) = 49.47$, falls short of the observed flux by a factor of 7. This suggests that a cluster of young massive stars is responsible for the ionization of the H\,{\sc ii} region, although the stellar population of this prominent source remains poorly characterized. \citet{Nogueras-Lara_2024} pointed out that there is an extensive population of young stars distributed throughout the Sgr~C region with ages $\ge 20$~Myr; these are old enough that the most massive remaining stars (B-type) are unlikely to provide the UV flux required to account for the Sgr~C H\,{\sc ii} region.

There is one notable exception: in their compilation of massive stars in the CMZ, \citet{Clark_et_al_2021} list two dusty Wolf-Rayet (WR) stars whose positions on the sky closely coincide with the Sgr~C H\,{\sc ii} region. WR stars of the WC subgroup are characterized by strong emission lines of ionized carbon, indicating high rates of mass and momentum loss. On average, WCL stars can deliver a few times $10^{49}$~s$^{-1}$ for $N_{{\rm LyC}}$ \citep[e.g.,][]{Figer_et_al_1999, Martins_et_al_2007}. The positions of both stars are superposed on our Fig.~\ref{ZoomHII}: object J174437 coincides closely with the center of the mid-IR energy distribution observed with SOFIA/FORCAST \citep{Hankins_et_al_2020}, while J174440 is slightly offset toward the [CII] "bubble". We considered the possibility that the driving sources could have moved from their formation sites, but given the young dynamical age of the shell (0.13~Myr), significant migration is unlikely. Both stars are classified as dusty, late WC-type (WCLd). Their spectral features are broad \citep[see Fig.~4 in][]{Geballe_et_al_2019}, but unfortunately, the low spectral resolution of the data does not allow for the derivation of detailed kinematic information. Because WC stars outside the Arches and GC clusters are rare, and because of their close morphological correspondence, we follow the assumption that both objects are physically associated with Sgr~C.

To what extent the two known WR stars in Sgr~C are able to contribute to the Lyman continuum flux required to maintain the H\,{\sc ii} region is difficult to address. On average, WCL stars may deliver a few times $10^{49}$~s$^{-1}$ for $N_{{\rm LyC}}$ \citep[e.g.,][for the Quintuplet cluster]{Figer_et_al_1999}, but for dusty types this may not apply.

Beyond this consideration, the mere existence of these objects illuminates the star formation history in Sgr~C. WC stars represent the advanced evolutionary stage of massive stars ($> 18~M_{\odot}$ initial mass; \citealt{Sander_et_al_2019}) prior to core collapse. The WC phase is relatively short (a few 100~kyr) compared to the lifetime of the progenitor, which means the star formation event must have happened within the last several million years. The fact that the surprisingly short kinematic age of the [CII] shell (0.13~Myr; Table~\ref{shell_properties}) is comparable to the duration of the WC phase may be taken as evidence that one or both of the WC stars have had a significant impact on the evolution of the shell. We investigate their possible mechanical feedback into the surrounding gas in the following sections.

\subsubsection{Thermal expansion of the HII-region}
The thermal expansion of an H\,{\sc ii} region into a uniform medium has been described by \citet{Spitzer_1978}. In the absence of radiation pressure, the temporal evolution of a D-type ionization front (IF) is described by 
\begin{equation}
R_{\rm IF} = R_{\rm St} \, \left[1 + \frac{7}{4} \frac{c_{\rm i} t}{R_{\rm St}}\right]^{4/7}
\end{equation}
with 
\begin{equation} 
R_{\rm St} \, [{\rm pc}] = 1.3 \, \left(\frac{N_{\rm Lyc}}{10^{50}~{\rm s}^{-1}}\right)^{1/3} \left(\frac{n_0}{10^3~{\rm cm}^{-3}}\right)^{-2/3} \left(\frac{T}{6800~{\rm K}}\right)^{0.28}, 
\end{equation}

where $R_{\rm St}$ is the classical Str\"omgren radius in convenient units \citep[Eq.~15.3;][]{Draine_2011}, and $c_{\rm i} \approx 11 \sqrt{T/6800~{\rm K}}$~\kms\ is the isothermal sound speed in the ionized gas. For the Sgr~C parameters, we calculate $R_{\rm St} \sim 1.3$~pc. The observed shell radius ($R_{\rm IF} \approx 3$~pc) would have been reached after $2.6 \times 10^5$~yr. However, by that time, the front would have slowed down to $\sim 0.5 c_{\rm i}$, which is obviously incompatible with the observed, much faster shell expansion.

\subsubsection{Are stellar winds driving the shell?}

The WC phase is characterized by high rates of mass and momentum loss. \citet{Sander_et_al_2019} report mean parameters for WC8-type stars of $V_{\text{wc}} = 1810$~\kms\ (terminal wind velocity), a mass-loss rate ($\dot{M}$) of $3 \times 10^{-5}$~$M_{\odot}~\text{yr}^{-1}$, and a luminosity of $4 \times 10^5$~$L_{\odot}$. Detailed investigations of the stellar population of the Quintuplet cluster reveal large variations in these parameters, with values of $2$--$10 \times 10^{-5}$~$M_{\odot}~\text{yr}^{-1}$ for WCL stars, most of which are identified as "dusty" members of binaries \citep{Gallego-Calvente_et_al_2022}. The current momentum input of that cluster is $0.78$~$M_{\odot}~\text{yr}^{-1}$~\kms, dominated by its WCL stars. Analysis of the slightly younger Arches cluster \citep{Gallego-Calvente_et_al_2021} yields a similar value, $1.2$~$M_{\odot}~\text{yr}^{-1}$~\kms.

To return to Sgr~C, we investigated whether the observed [CII] shell has been swept up by stellar wind(s). In view of the current broken-up, fragmented appearance of the shell, its development is likely better described by a momentum-conserving approach. The (assumed constant) momentum input is then given by \citep[e.g.,][and references therein]{Sakashita_et_al_1984}:

\begin{equation}
\dot{M} \cdot V_{w} = \frac{8\pi}{3} (R_{sh}V_{sh})^2 \rho_{0}.
\end{equation} 
 The density of the ambient gas, $\rho_0$, is estimated from the accumulated shell mass assuming a uniform distribution (the equivalent $\text{H}_2$ particle density is $0.7$--$3.7 \times 10^3$~cm$^{-3}$). Using the parameters from Table~\ref{shell_properties}, we calculate the required momentum input to be $1.6$--$8.2$~$M_{\odot}$~yr$^{-1}$~\kms, which exceeds the momentum rate potentially provided by a typical WCL-type star ($\sim 5 \times 10^{-2}$~$M_{\odot}$~yr$^{-1}$~\kms) by a factor of 100. Taking radiation pressure into account does not ease this discrepancy, as in WC-type stars specifically, the momentum transfer to the gas is dominated by the wind ($\dot{M} V_{\text{w}} \gg L_{\star}/c$; e.g., \citealt{Nugis_2000}). This mismatch indicates that even the integrated stellar wind feedback from star clusters an order of magnitude more (UV-)luminous than Sgr~C falls short of the required momentum. Even given the inherent uncertainties in our simplifying assumptions, it remains difficult to reconcile these disparate numbers. 
 
This results in a challenging puzzle: the short kinematic age of the expanding shell suggests an acceleration during the last hundred thousand years, yet the only identified stars (the two WCLd) cannot provide the necessary momentum to drive the shell during their short Wolf-Rayet phase. If the expansion was instead driven by less powerful winds during the few million years of main-sequence evolution of the progenitor stars (or other, as-yet-undiscovered cluster members), the mean expansion rate would have been slower to reach the observed radius. With a better characterization of the Wolf-Rayet stars, one could quantify a scenario involving a later, more recent acceleration of the shell due to their stronger winds expanding into a pre-existing bubble. Prior to breakup, the adiabatic expansion of the trapped bubble would have been more efficient. We note that \citet{Tiwari_et_al_2021}, in their discussion of the dynamics of the wind-driven shell RCW~49 (which has comparable shell parameters), encounter similar contradictory numbers. Unlike RCW~49, where the embedded compact cluster, Westerlund~2, has been studied in great detail, for Sgr~C we lack constraining information regarding the associated stellar population. A dedicated search for additional cluster members, and in particular, high-resolution spectroscopy of the identified WCL stars to refine their feedback parameters, would provide valuable input to this discussion. Candidate stellar cluster members have been revealed in the recent 25 and 37~$\mu$m SOFIA/FORCAST survey of the region \citep{Zhao_Sub_2025, Cotera+24}. 

In parallel, this analysis would benefit from further constraints on the physical characteristics of the gas in the Sgr~C shell, including $\text{C}^+$ excitation and the derived mass. While we have attributed the kinematics of the shell to expansion alone, the observed profiles may be partially influenced by the ablation of gas from the inner edge of the shell, as suggested by \citet{Bonne_et_al_2023} in their analysis of the shell associated with RCW~79. Improved [CII] spectra with higher S/N from a dedicated follow-up project would provide better constraints on the gas kinematics. Without SOFIA, alternative tracers accessible from ground-based observatories must be utilized. A suitable dense-gas PDR tracer would be, for example, the $J=6$--$5$ rotational transition of CO, which, due to its higher excitation, should not be affected by line-of-sight confusion. The kinematics of the ionized gas in the bubbles could be further investigated using hydrogen recombination lines. High angular and spectral resolution (similar to our [CII] survey) to resolve the shell kinematics is mandatory\footnote{The only published high-angular-resolution recombination line spectrum known to us, the VLA H$72\alpha$ line presented by \citet{Liszt_Spiker_1995} in their Fig.~3, was extracted close to our position \#13 (Fig.~\ref{ZoomHII}), and its non-Gaussian shape resembles the kinematics reflected in the [CII] profile.}.

\subsubsection{Buried supernova explosion?}
The momentum stored in the expanding shell is on the same order of magnitude as that of a supernova remnant in its later, momentum-conserving snowplow phase ($M_{\text{shell}} \gg M_{\text{ejecta}}$). Observable evidence of an embedded supernova explosion releasing a typical energy of $10^{51}$~erg approximately 10~kyr ago into a dense ($10^3$~cm$^{-3}$) environment would have long since faded away. Several studies \citep[e.g.,][]{Shull_1980, Wheeler_et_al_1980} have examined the enhanced rate of evolution of such "buried" supernovae, where the cavity and the expanding shell of swept-up material remain the primary signs of the event until the remnant merges with or breaks out of the ambient cloud. For expansion into a uniform medium of density $10^3$~cm$^{-3}$, the Sedov--Taylor phase is short; after a few thousand years, the shell evolves through a pressure-driven phase—defined by efficient cooling of the hot gas—into the momentum-conserved phase \citep[see][and references therein]{Kim_Ostriker_2015, Haid_et_al_2016}.

At this stage, the momentum transferred to the expanding shell is described by $p_{\text{mc}} \sim 10^5 (n/10^3)^{-0.16}$~$M_{\odot}$~\kms\ (for an injected energy of $10^{51}$~erg). This is consistent with the analytical description of the momentum-driven snowplow by \citet{Wheeler_et_al_1980}, which yields $p_{\text{mc}} \sim 1.5 \times 10^5$~$M_{\odot}$~\kms. The shell radius develops as $R_{\text{sh}} \propto t^{1/4}$ during this phase. From the observed properties, we estimate the dynamical age of the remnant to be $\sim 2$--$3 \times 10^4$~yr, assuming an expansion velocity of $150$--$200$~\kms\ at the transition from the pressure-driven to the momentum-driven phase \citep[reflecting the steep increase in the cooling function with decreasing shock velocity and post-shock temperature;][]{Draine_2011}. The luminosity behind this relatively slow shock would be low, $\sim 2 \times 10^4$~$L_{\odot}$, with a temperature of only a few $10^4$~K.

While an embedded supernova remnant in this late stage is no longer expected to emit strong X-rays, there is a relatively weak local X-ray peak coincident with the Sgr~C H\,{\sc ii} region (Fig.~\ref{shell_tracers}; using X-ray data from \citealt{Wang_2021}). The low X-ray intensity of this peak might be at least partially attributed to extinction by the dense foreground cloud to the southeast, which contains the nearby EGO where protostars and YSOs are currently forming \citep{Kendrew_et_al_2013, Lu_et_al_2021, Crowe_et_al_2025}. This peak might also be associated with a brighter patch of X-ray emission located $0.04^\circ$ to the southwest of the H\,{\sc ii} region, which could be an unrelated line-of-sight coincidence. The X-ray emission from these two peaks and an extension to the north was originally reported by \citet{Tsuru_et_al_2009}, who found that the spectrum and overall luminosity were consistent with a typical supernova remnant. However, because their relatively low-resolution observations with the \textit{Suzaku} satellite did not resolve the two peaks, and because the centroid of their X-ray source was displaced by $2'$ (5~pc) from the Sgr~C H\,{\sc ii} region, \citet{Tsuru_et_al_2009} concluded that the source was a new supernova remnant candidate unrelated to the Sgr~C H\,{\sc ii} region. Nevertheless, the possibility that the weak source component coincident with the Sgr~C H\,{\sc ii} region (shown in Fig.~\ref{shell_tracers}) is the last vestige of an old supernova remnant cannot be entirely ruled out.

An alternative to the supernova scenario is that the X-rays are produced by a strong shock where the high-velocity winds from the WR stars encounter the inner edge of the expanding shell. However, given the significant momentum discrepancy discussed previously, the supernova remnant scenario remains a compelling explanation for the shell's dynamics. Further, more sensitive X-ray observations that allow for higher spatial resolution and better spectroscopic constraints are needed to investigate these possibilities (see for e.g, Zhu et al, Submitted).

\subsection{Are Sgr~C winds driving the nonthermal filament?} \label{gas_NTF}

\citet{Liszt_Spiker_1995} emphasized the correspondence of the eastern boundary of the H\,{\sc ii} region with the prominent, long, nonthermal radio filament. We argued in Sect.~\ref{sec:shell-kinematics} that the morphology and kinematics of [CII] suggest that the Sgr~C shell has broken up along its eastern perimeter. We further investigate this scenario in Fig.~11, where we compare the [CII] emission integrated from $-45$ to $-50$~\kms\ with the 1.28~GHz MeerKAT intensity and spectral index data \citep{Heywood_et_al_2019_nature, Heywood_et_al_2022}. Strikingly, the NTF skirts the edge of the [CII] shell, runs through the periphery of the ionized gas bubble manifested in radio continuum emission, and is superimposed on the center of the apparent adjoining [CII] bubble, suggesting a causal connection between the NTF and the H\,{\sc ii} region and its [CII] entourage. This geometrical relationship could provide an important clue to the origin of the NTF. 

Indeed, there are at least two possible reasons why H\,{\sc ii} regions in the Galactic center might be a source of relativistic electrons that can diffuse along the local magnetic field lines and illuminate the local field with their synchrotron emission, producing filamentary structures aligned with the magnetic field.

The first, proposed by \citet{Rosner_Bodo_1996}, invokes diffusive shock acceleration of electrons to relativistic energies in the terminal shocks of stellar winds. This model was extended to the interaction of the collective winds of massive stars in Galactic center clusters by \citet{Yusef-Zadeh_2003}. This model can be well accommodated by our source model for Sgr~C. The collective winds from the presumed young star cluster powering the Sgr~C H\,{\sc ii} region will expand until they encounter the denser, swept-up gas defining the [CII] shell, where they undergo the shocks in which the relativistic electrons are produced. Those electrons are injected into the local magnetic field near their point of acceleration and are constrained by the Lorentz force to diffuse along the local magnetic field lines, illuminating those field lines with their synchrotron emission. However, \citet{Rosner_Bodo_1996} argue that the transverse dimension of an NTF produced in this way is set by the radius of the wind bubble, which is not the case for the H\,{\sc ii} region or [CII] shell in Sgr~C. The width of the NTF is much smaller than the size of those features, so perhaps the conditions favorable for the requisite shock acceleration only take place on the eastern side of the [CII] shell, where it appears to break out into the adjoining bubble.  

The second potential source of relativistic electrons, suggested by \citet{SerabynMorris94}, invokes magnetic field line reconnection at the ionized surfaces of molecular clouds in the CMZ as the mechanism that accelerates electrons. This can occur because the magnetic field orientation within the dense clouds of the CMZ is largely oriented close to parallel to the Galactic plane \citep{Chuss+03, Pare+24}, presumably as a result of tidal shear in the CMZ \citep{Novak+2000}. On the other hand, the magnetic field geometry in the intercloud medium within the central few hundred parsecs appears to be predominantly perpendicular to the Galactic plane, as evidenced by the fact that the abundant nonthermal radio filaments in the central few hundred parsecs of the Galaxy have a strong propensity to be perpendicular to the Galactic plane \citep[e.g.,][and references therein]{Heywood_et_al_2022}. Consequently, at the ionized surface of a molecular cloud where electrons are abundant, and where the two roughly orthogonal magnetic field systems can interact, reconnection can accelerate the electrons and inject them into the ambient local field, producing a nonthermal radio filament. This scenario was invoked by \citet{Uchida_et_al_1996} for the origin of the "Snake" NTF, which, like the Sgr~C NTF, apparently originates in an H\,{\sc ii} region (which, unfortunately, was outside our [CII] survey).   

Whatever acceleration mechanism produces the relativistic electrons, they should be able to diffuse in both directions along continuous magnetic field lines. However, the Sgr~C NTF is quite short to the south compared to its large extent to the north. As can be seen in Figs.~5, 15, and 8, the Sgr~C molecular cloud is located to the south of the H\,{\sc ii} region, at least in projection, so we propose that the cloud is blocking much of the southern-directed flow of relativistic electrons. It also appears that the field lines illuminated by synchrotron radiation are bent slightly to the west where they are first superimposed upon the molecular cloud (Figs.~15 and 6), and the emission from the NTF becomes much weaker beyond the bend to the south. Consequently, we suggest that many of the emitting electrons are blocked or reflected where the illuminated filament meets the molecular cloud, but some survive to faintly illuminate the magnetic field beyond the point where the cloud has deformed the field.

\subsection{Implications for star formation in Sgr~C}

Previous observations suggest that star formation in the molecular cloud associated with Sgr~C is a persistent process spanning several orbital periods of the source around the Galactic center. An analysis of the stellar population toward Sgr~C by \citet{Nogueras-Lara_2024} reports evidence for a massive star formation event 20~Myr ago and an intermediate-age stellar population with ages between 2 and 7~Gyr. The stars powering the Sgr~C H\,{\sc ii} region and/or driving the expanding shell discussed here likely formed during a more recent event, several Myr ago. 

Currently, several sites of active massive star formation have been identified (as reviewed in Sect.~1). Their location at the inner edge of the [CII] shell, toward the tip of the ambient molecular cloud closest to the H\,{\sc ii} region in projection (Figs.~5 and 6), is very suggestive of star formation being triggered by the expanding shell. However, interestingly, given the short dynamical age of the shell, star formation must have proceeded rapidly after the passage of the shock. This scenario is perhaps analogous to that of the prototypical bubble RCW~120, for which \citet{Luisi_et_al_2021} proposed that stellar feedback has triggered star formation. It is also possible that pre-existing star formation activities formed the shell as a consequence of their early feedback; the shell's expansion likely accelerated or triggered further star formation in adjacent regions. 

The most obvious requirement to be fulfilled is that the dynamical time, $t_{\text{dyn}}$, must exceed the free-fall time, $t_{\text{ff}} = 0.13 \times (10^{5}~\text{cm}^{-3}/n(\text{H}))^{1/2}$~Myr, where $t_{\text{ff}}$ is the time for a uniform object of density $n(\text{H})$ to collapse under its own gravity. With $t_{\text{dyn}} \sim 0.13$~Myr, we calculate the lower limit to the core's pre-collapse density to be $n(\text{H}) > 10^{5}$~cm$^{-3}$. This is high, but still a rather comfortable constraint for a fast-moving shell running into a medium-dense ($10^3$~cm$^{-3}$) and likely clumpy ambient cloud. At these high densities, cooling of the shocked, swept-up layer is efficient, and shock compression in the limiting case of an isothermal shock ($T \sim 100$~K) will be high, roughly proportional to $(V_{\text{sh}}/C_{\text{m}})^2$ \citep{Draine_2011}. Given the sound velocity in the molecular gas $C_{\text{m}} \sim 0.6$~\kms, the gas can be compressed by three orders of magnitude. 

Considering that the typical evolutionary age of an Extended Green Object (EGO) is approximately $10^3$~years \citep[e.g.,][]{Chen_et_al_2013}, the triggering of star formation in the EGO by the expanding [CII] shell is a compelling scenario warranting further exploration. 

\section{Summary and outlook}
We imaged the [CII] fine-structure transition across the central 74$\times$47~pc of the Sgr~C complex using the upGREAT receiver at the SOFIA telescope. This study provides a detailed look at this massive area within the CMZ. Through high-resolution spectroscopic imaging of the [CII] line, supported by APEX CO observations, we probed the complex dynamics and morphology of the interstellar medium, revealing and exploring the widespread and coherent emission stretching from the Sgr~A to Sgr~C regions.

Our work identified several distinct emission components, including the H\,{\sc ii} regions FIR~4 and Source~C. We focused here on the most prominent feature: the expanding shell associated with the Sgr~C H\,{\sc ii} region. Based on the lack of recombination line emission at the shell velocities, we identify this feature as [CII] originating from a PDR. Our main findings regarding this feature are:

 \begin{enumerate} 
    \item \textbf{Kinematics and Age:} A detailed kinematic analysis of the shell yields a high expansion velocity of 23.4~\kms\ and a correspondingly short dynamical age of only 0.13~Myr.
    
    \item \textbf{Mass and Energy:} We calculated the shell's physical properties, estimating a total gas mass of $\sim 1.9 \times 10^4~M_{\odot}$ and a kinetic energy of $\sim 10^{50}$~erg. These values are conservative lower limits due to opacity effects.
    
    \item \textbf{Driving Mechanism:} A central finding of our work is that the stellar winds from the known massive stars within the region, including two recently identified Wolf-Rayet stars, appear to be insufficient to power the observed expansion. The calculated momentum of the shell exceeds the integrated wind momenta of the known stars by nearly a factor of 100. This discrepancy suggests that alternative drivers, such as a past "buried" supernova explosion, are required to explain the observations.
    
    \item \textbf{Nonthermal Filament Connection:} We find a striking spatial association between the expanding shell and the bright Sgr~C nonthermal radio filament (NTF). The NTF appears to skirt the edge of the [CII] shell, suggesting a physical interaction where the collective stellar winds or shocks associated with the shell could provide the relativistic electrons for the filament via diffusive shock acceleration.
    
    \item \textbf{Triggered Star Formation:} The location of active high-mass star formation (the EGO) at the edge of the expanding shell is consistent with a triggered star formation scenario. Our calculations show that the high density of the pre-collapse core allows for a free-fall time short enough to be consistent with the young dynamical age of the shell.
 \end{enumerate} 

We conclude that future observations, such as high-spatial-resolution radio recombination line observations, additional dense gas PDR tracers (e.g., high-$J$ CO lines), and further, more sensitive X-ray observations that allow for higher spatial resolution and better spectroscopic constraints \citep[e.g.,][]{Zhu_et_al_2026} are needed to determine the precise nature of the driving force. The [CII] data from this study are available from the SOFIA archive at \url{https://irsa.ipac.caltech.edu/}.

\begin{acknowledgements}
This work is based on observations made with the NASA/ DLR Stratospheric Observatory for Infrared Astronomy (SOFIA). SOFIA was jointly operated by the Universities Space Research Association, Inc. (USRA), under NASA contract NNA17BF53C, and the Deutsches SOFIA Institut (DSI) under DLR contract 50 OK 0901 to the University of Stuttgart.
Financial support for this work was provided by NASA through awards 05--0022 and 06--0173 issued by USRA, by the Max-Planck-Institut f\"ur Radioastronomie, and by the Deutsche Forschungsgemeinschaft (DFG) through the SFB 956 program award to MPIfR and the Universit\"at zu K\"oln. 
This research made use of Spitzer data from the NASA/IPAC Infrared Science Archive, which is operated by the Jet Propulsion Laboratory, California Institute of Technology, under contract with the National Aeronautics and Space Administration; data from the Herschel Science Archive, which is maintained by ESA at the European Space Astronomy Centre; and data products from the Midcourse Space Experiment, whose data processing was funded by the Ballistic Missile Defense Organization with additional support from NASA Office of Space Science.
D. Riquelme-V\'asquez acknowledges the financial support of DIDULS/ULS, through the project PAAI 2023.
R. Simon gratefully acknowledges the Collaborative Research Center 1601
(SFB 1601 subproject B2) funded by the Deutsche Forschungsgemeinschaft
(DFG, German Research Foundation) -- 500700252. We thank the anonymous referee for critical reading and constructive comments that helped to improve this manuscript. 
This research made use of Astropy, a community-developed core Python package for Astronomy \citep{Astropy_2013}.          

\end{acknowledgements}

\clearpage
\bibliographystyle{aa} % style aa.bst
\bibliography{references_CMZ} % your references Yourfile.bib

\appendix
\section{Calculation of the C$^+$ Column Density}
\label{sec:columndensity}

The observed line intensity of the [CII] 158 $\mu$m  fine-structure transition depends on the local physical conditions \citep[see ][for thorough discussions]{Crawford_et_al_1985, Goldsmith_et_al_2012}, means the nature of the collision partners (H$_2$, HI or e$^-$), the density (n) and kinetic temperature (T$_{kin}$) of the emitting gas. Also the optical depths of the transition needs to be constrained in order to reliably calculate the  C$^+$ column density, N(C$^+$), from the observed velocity-integrated line intensity, $\int$T$_b$ dV [K km\,s$^{-1}$]. \\
In Sect.3.2.1 we have argued that the [CII] spectra observed towards Sgr\,C can be decomposed into emission from both, the neutral gas in the Photodissociation Region (the shell) and the ionized HII region. 
We will discuss the C$^+$ excitation in the PDR first.

In the limiting case of optically thin emission and assuming no background radiation (Eq.26 of Goldsmith et al.) N(C$^+$) [cm$^{-2}$] is given by 
\begin{equation}
N(C^+)\ = 2.91\times 10^{15} \times \Bigg 
[1 + 0.5\ e^{\frac{91.25}{T_{kin}}}\Bigg( 1 + \frac{2.4\times 10^{-6}}{n\ R_{ul}}\Bigg )\Bigg] \ \int{T_b \  d V},
\end{equation}
where R$_{ul}$(T$_{kin}$) is the downward collisional rate coefficient. As we have no other constraining observations, for the Sgr C shell we adopt typical numbers derived for galactic PDRs (reference), namely (n $\sim$10$^{3}$ cm$^{-3}$, T$_{kin}$$\sim$100 K). Assuming next that the main collision partner is molecular hydrogen, from Eq. 43 in \citep{Goldsmith_et_al_2012}, R$_{ul}$(T$_{kin}=100$ K$)=3.8\times10^{-10}$ cm$^3$\,s, we obtain 
\begin{equation}
N_{PDR}(C^+) \ [cm^{-2}] = 2.93 \ 10^{16} \ \eta_{\tau} \ \int{T_b \ d V}
\end{equation}
To quantify uncertainties, the pre-factor does change by factor of 2 for H$_2$ densities of 0.5 and 3$\times$10$^3$ cm$^{-3}$, respectively.
While this conversion only holds in the optically thin limit ($\eta_{\tau}$ =1), this is an unlikely scenario for the prominent Sgr C PDR. Unfortunately, due to the heavy velocity-crowding along the line-of-sight towards the galactic center, constraining the [$^{12}$CII] line opacity by observations of the isotopic [$^{13}$CII] triplet \citep{Guevara_et_al_2020} has been unsuccessful towards Sgr C. To quantify the underestimation of the column density, we run RADEX, the non-LTE radiative transfer code provided by \citet{vanderTak_et_al_2007}, for a line opacity $\tau$ $\sim$2, typical on larger scales for galactic PDRs \citep{Guevara_et_al_2020}. For the observed line intensities, the correction factor $\eta_{\tau}$ calculates to 1.5 (for 10$^{3}$ cm$^{-3}$) and 1.8 (3$\times$10$^{3}$ cm$^{-3}$, 100 K in both cases). Hence, in the following, we calculate the C$^+$ column density across the PDR from 
\begin{equation}
N_{PDR}(C^+) \ [cm^{-2}] = 3.5 (\pm 1.0) \times 10^{16} \ \int{T_b \ d V}, 
\end{equation}
where the uncertainty reflects the above variation in density only. \\
We note, because the assumption is often made in the analysis of [CII] data, that for optically thin, thermalized (n >> n$_{cr}$ $\sim$6000 cm$^{-3}$, for collisions with H$_2$ at 100 K) emission, factor 5 lower C$^+$ column densities would be derived
\begin{equation}
N_{ot,t}(C^+) \ [cm^{-2}] = 6.5 \times 10^{15} \ \int{T_b \ d V}, 
\end{equation}

Similarly, we can calculate the C$^+$ column density for ionised gas residing in the HII-region. 
In an equivalent formulation of Equation A.1 (e.g. \citep{Langer_et_al_2014})
\begin{equation}
N(C^+)\ = 2.91\times 10^{15} \times \Bigg 
[1 + 0.5\ e^{\frac{91.25}{T_{kin}}}\Bigg( 1 + \frac{n_{cr}}{n_e})\Bigg] \ \int{T_b \  d V}.
\end{equation}
For the electron temperature range T$_e\sim$ 6-8000 K, n$_{cr}$ $\sim$ 35-44 cm$^{-3}$ (\citep{Goldsmith_et_al_2012}, Table 2). With n$_e$ $\sim$120$\pm13$ cm$^{-3}$ (from the radio flux and electron temperature discussed above) we calculate
\begin{equation}
N_{HII}(C^+) \ [cm^{-2}] = 4.87 (\pm 0.1) \times 10^{15} \ \int{T_b \ d V}.
\end{equation}

The total hydrogen column density, N(H), is then estimated from N(C$^+$), assuming all carbon is single-ionized and for a fractional abundance C/H $\le$ 3 $\times$ 10$^{-4}$ \citep{Genzel_et_al_1990}.

\begin{equation}
N(H) \ [cm^{-2}] = 1.17 (\pm 0.3) \ 10^{20} \ \int{T_b \ d V} \ [K\ km\,s^{-1}]. 
\end{equation}

\section{Complementary figures}
\label{complementaryplots}
\begin{figure*}
\centering
\includegraphics[width=\hsize, angle=0]{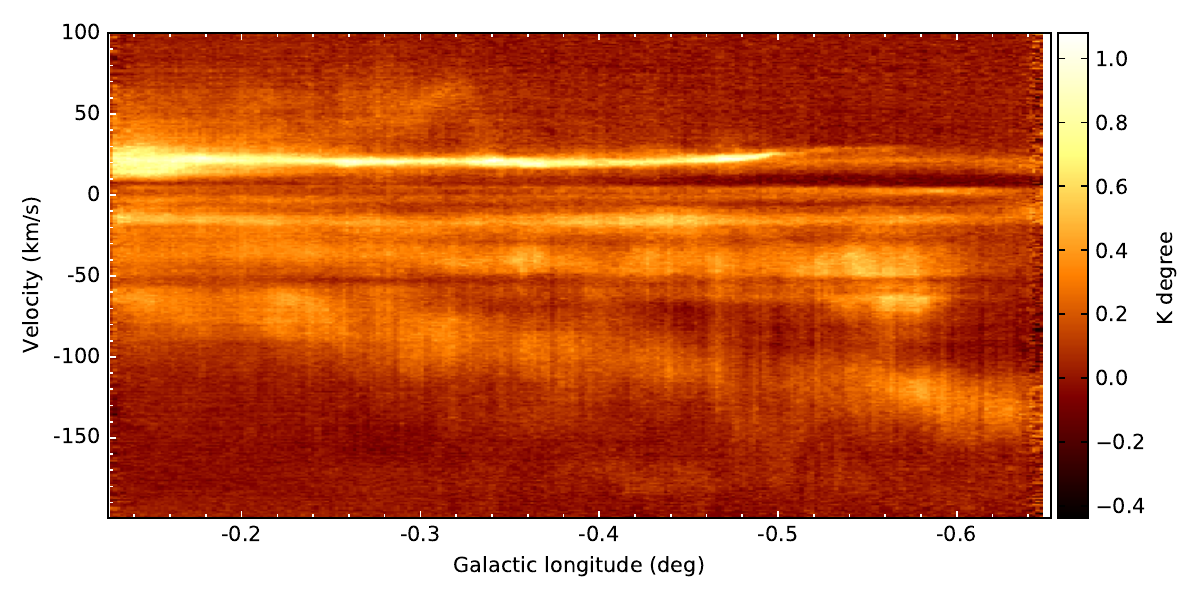}
\caption{[CII] longitude-velocity diagram integrated over the observed
  latitude range (from $-0\deg.21$ to $0\deg.12$). The diagonal stripe spanning $-70$ to $-130$ \kms, likely belongs to the "100 pc stream" \citep{Kruijssen_et_al_2015}.}
\label{lvdiagram}
\end{figure*}

\begin{figure*}
\centering
\includegraphics[width=0.8\hsize, angle=0]{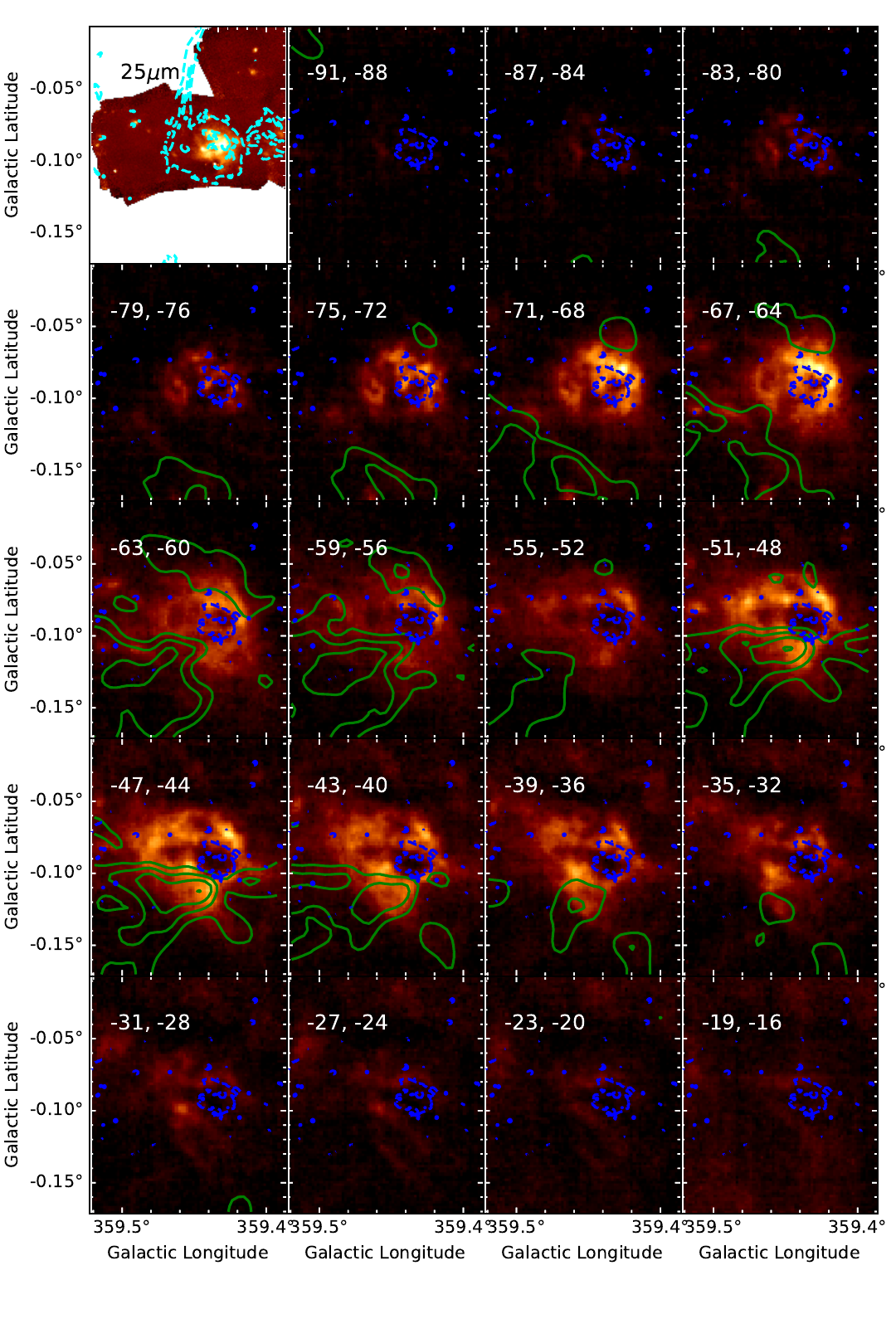}
\vspace{-8mm}
\caption{Zoom towards the HII region in [CII]. The first panel on the
  left serves as a reference, displaying the FORCAST 25 $\mu$m emission
  \citep{DeBuizer_et_al_2025, FORCAST_data} in a color scale, while the dashed cyan
  contours represent the 10\%, 50\%, and 90% intensity levels of the 
  MeerKAT radio continuum image centered at 1.28 GHz \citep{Heywood_et_al_2022}. Moving to the subsequent
  panels, we showcase the integrated intensity of the [CII] emission
  in increments of 4 km\,s$^{-1}$. The color scale is linear from 0 to
  60 K km\,s$^{-1}$. The velocity range is indicated in the upper left corner
  of each plot. The blue dashed contours denote the 20\% and 60\%
  intensity levels of the FORCAST 25 $\mu$m emission, while the green
  contours depict 30\%, 50\%, 70\%, and 90\% of the intensity peak (29
  K km\,s$^{-1}$) of $^{13}$CO (2-1) channel maps within the same velocity
  range as the [CII] emission.}
\label{zoomHIIregion_channel1}
\end{figure*}

\end{document}